\RequirePackage{fix-cm}
\RequirePackage{amsmath,amssymb,amsfonts} 
\documentclass[smallextended]{svjour3}       
\smartqed
\usepackage{algorithmic} 
\usepackage{textcomp} 
\usepackage{xcolor} 
\usepackage{float} 
\usepackage[T1]{fontenc}
\usepackage{flushend}
\usepackage{soul}
\usepackage{graphicx}
\usepackage[caption=false]{subfig} 
\usepackage{natbib}

\usepackage{booktabs}
\usepackage{multirow}

\usepackage{colortbl}
\definecolor{Gray}{gray}{0.88}
\newcolumntype{a}{>{\columncolor{Gray}}c}

\usepackage{tcolorbox}
\usepackage{url}
\usepackage[ruled,vlined]{algorithm2e}
\newcommand{\tildex}[1]{$_{\widetilde{~}}$#1}
\newif\ifdiff
\newcommand{\revisionone}[1]{\ifdiff{\leavevmode\color{blue}{#1}}\else{#1}\fi}
\newcommand{\revisiontwo}[1]{\ifdiff{\leavevmode\color{red}{#1}}\else{#1}\fi}
\difffalse

\begin{document}

\title{\textit{Will You Come Back to Contribute?} Investigating the Inactivity of OSS Core Developers in GitHub}

\titlerunning{Will You Come Back to Contribute?}        

\author{Fabio Calefato        \and
        Marco Aurelio Gerosa \and
        Giuseppe Iaffaldano \and
        Filippo Lanubile \and
        Igor Steinmacher
}

\authorrunning{F. Calefato et al.} 

\institute{F. Calefato \at
              University of Bari, Italy \\
              \email{fabio.calefato@uniba.it}           
           \and
           M.A. Gerosa \at
              Northern Arizona University \\
              \email{marco.gerosa@nau.edu}
            \and
           G. Iaffaldano \at
               University of Bari, Italy \\
              \email{giuseppe.iaffaldano@uniba.it}
            \and
           F. Lanubile \at
               University of Bari, Italy \\
              \email{filippo.lanubile@uniba.it}
            \and
           I. Steinmacher \at
              Northern Arizona University  \\
              \email{igor.steinmacher@nau.edu}
}

\date{Received: date / Accepted: date}

\maketitle

\begin{abstract}
Several Open Source Software (OSS) projects depend on the continuity of their development communities to remain sustainable. Understanding how developers become inactive or why they take breaks can help communities prevent abandonment and incentivize developers to come back. In this paper, we propose a novel method to identify developers' inactive periods by analyzing the individual rhythm of contributions to the projects. Using this method, we quantitatively analyze the inactivity of core developers in 18 OSS organizations hosted on GitHub. We also survey core developers to receive their feedback about the identified breaks and transitions. 
Our results show that our method was effective for identifying developers' breaks. About 94\% of the surveyed core developers agreed with our state model of inactivity; 71\% and 79\% of them acknowledged their breaks and state transition, respectively. We also show that all core developers take breaks (at least once) and about a half of them (\tildex{45\%}) have completely disengaged from a project for at least one year.
We also analyzed the probability of transitions to/from inactivity and found that developers who pause their activity have a \tildex{35} to \tildex{55}\% chance to return to an active state; yet, if the break lasts for a year or longer, then the probability of resuming activities drops to \tildex{21-26}\%, with a \tildex{54}\% chance of complete disengagement. These results may support the creation of policies and mechanisms to make OSS community managers aware of breaks and potential project abandonment.
\end{abstract}


\section{Introduction}

The success of an OSS project depends on the strength and health of the group behind it~\citep{Crowston.Howison.2016, Link.Gemonprez.2018, coelho2017modern}. \cite{marsan2018} reported that, from the perspective of OSS developers, maintainers, and managers, the loss of contributors is a major problem in OSS communities and ecosystems, followed by poor code quality. Almost half of the OSS projects that fail do so because of a problem related to the development team~\citep{coelho2017modern}. Turnovers often disrupt the community, reduce productivity~\citep{mockus2010}, and degrade product quality~\citep{schilling2014we,foucault2015}, especially when core contributors are involved. More specifically, \cite{avelino2019} showed that 59\% of the analyzed projects did not survive after the departure of \textit{Truck Factor} developers~\citep{williams2002pair}, i.e., developers who hold unique responsibilities in the project. 

Therefore, it is critical to nurture the project workforce and avoid losing core developers and their knowledge. However, little is known about either the stages of project disengagement that developers undergo, or the transitions between these stages. So far, research has focused on the developers' life cycle: how people join the projects~\citep{krogh.ea:2003}, including the barriers they face~\citep{SteinmacherCSCW.ea:2015,Hannebauer:2017,balali2018newcomers}; how they are attracted~\citep{yamashita.ea:2016, santos.ea:2013, Fronchetti.ea.2019}; and how they become long-term contributors or core members~\citep{ducheneaut2005, Nakakoji.ea:2002, zhou2012}. The limited research about understanding developers' disengagement has focused on the risks that projects incur when they lose developers~\citep{ricca2010heroes,avelino2019,ferreira2017comparison} and factors related to the developers' abandonment, using the survival analysis technique~\citep{lin2017developer}. To avoid developers' disengagement, researchers analyzed the factors related to developers' engagement and retention~\citep{schilling2014we,midha2007retention,zhou2016inflow}. \revisionone{However, these factors fail to consider the developers' rhythm or the disengagement stages and their transitions. 

In this paper, we employ a mixed-methods approach to investigate how developers take breaks and disengage from OSS projects. We first design a state model based on interviews with OSS developers. Then, we analyze data from eighteen OSS organizations and identify core developers' breaks by analyzing their personal work rhythm and identifying their inactivity states and the transitions between states. Then, we collect the developers' feedback about the output of our method. Finally, we employ our method to characterize break frequency and length as well as state transition probabilities in the analyzed OSS organizations. 

This paper makes the following contributions: (i) an empirically-defined state model that describes developers' (in)activity states and transitions in OSS projects; (ii) a method to identify the inactivity periods based on the individual rhythm of contribution, assessed by OSS developers via a survey; and (iii) an in-depth analysis at the organizational level about the frequency and extension of over 500 core developers' break periods, showing how the break periods vary and the probability of core developers' transitions to and from inactive periods. The proposed model captures disengagement, short breaks, and periods when contributors are only working on non-coding activities. Our results show that breaks are rather common, and highlight the importance of considering more than just coding to analyze inactivity/disengagement. The model and method proposed here are relevant for OSS communities; they can be used to define reference values for dashboards that alert when someone is likely about to leave.}

The rest of this paper is organized as follows. In Section~\ref{sec:rel_work}, we review related work. In Section~\ref{sec:res-framework}, we illustrate the research framework devised to carry out the study and answer the research questions. In Section~\ref{sec:phase_one}, we present the model developed to inform our research method, which is described in Section~\ref{sec:phase_two}. Results are reported in Section~\ref{sec:results} and discussed in Section~\ref{sec:discussion}. Finally, we conclude in Section~\ref{sec:conclusions}.

\section{Related Work}
\label{sec:rel_work}

In this section, we present the literature related to developers' life cycle and disengagement, focusing on OSS.

\subsection{Open Source Developers' Life Cycle}

Many OSS projects rely on a globally distributed community of developers who join projects for diverse reasons, such as reputation, giving back to the community, and learning~\citep{gerosa2021motivation,David.Shapiro:2008,Hars.Ou:2002,Lakhani.Wolf:2005,Krogh.Haefliger.ea:2012}. Many studies have focused on investigating the stages and activities in the path to becoming core members or long-term contributors. For example, \cite{Nakakoji.ea:2002} proposed the \textit{Onion model}, which represents the general structure of OSS roles that developers follow to become core members. \cite{krogh.ea:2003} proposed a joining script, based on steps followed to become a member of a project. \cite{ducheneaut2005} offered an in-depth look at a successful newcomer's socialization history, identifying activities that contributed to their success.

\cite{Steinmacher.ea:2014} bring a different perspective to the joining process, explaining it in two stages, namely onboarding and contributing, and describing the forces that push developers towards a project, such as motivation \citep{Hannebauer:2017, Krogh.Haefliger.ea:2012,gerosa2021motivation} and attractiveness \citep{yamashita.ea:2016, santos.ea:2013}, and those that hinder developers' onboarding \citep{SteinmacherCSCW.ea:2015}.

Overall, the studies that model OSS developers' life cycles address how to onboard, remain, and become long-term contributors. Even the motivation-related literature focuses on what compels developers to join projects~\citep{krogh.ea:2003, Hannebauer:2017, Hars.Ou:2002,silva2020google} and remain active~\citep{zhou2012, barcomb2014volunteer, yamashita.ea:2016}. Very little is known about developers' inactive periods and disengagement. Recent work has started filling this gap by investigating the impacts and reasons behind abandonment and turnover, as presented in the following section.

\subsection{Developers' Turnover and Disengagement}

Previous literature reported on the negative impact of turnover on team cognition and performance~\citep{levine2004,levine2005}, and its costs for the company and society~\citep{graef2000, garman2005}. Software engineering-specific literature has also shown that developers' turnover harms software development projects~\citep{bass2018employee, nakatsu2009comparative, hall2008}. \cite{hall2008} showed that projects with high turnover are less successful. \cite{mockus2010} reported that people leaving the organization (departures) might increase the probability of defects in commercial projects. Some researchers also investigate the impact of developers' turnover on software quality~\citep{mockus2009, foucault2015}. Regarding the reasons for leaving a software company, \cite{dittrich1985} and \cite{garden1988} reported that wage and pay raise equity is a significant predictor of intentions to quit. Additionally, \cite{lee2002social} showed that job satisfaction strongly influences turnover intentions.

In the OSS context, several researchers investigated turnover in OSS projects. For example, \cite{lin2017developer} studied why some developers are more likely to continue their contributions than others. Others focused on understanding the potential issues surrounding developers leaving OSS projects, including the so-called \textit{truck factor}~\citep{ricca2010heroes, avelino2016novel, ferreira2017comparison, cosentino2015}, which is defined as the number of people who have to be hit by a truck (i.e., leave the project) before the project itself is at risk \citep{williams2002pair}. \cite{avelino2019} found that truck factor is a real concern, which may affect project evolution. They showed that the majority of the projects do not survive when truck factor developers disengage and no other developers replace them.

Some other researchers have focused on understanding the reasons behind disengagement from OSS projects. \cite{Miller2019} found that contributors who work at night and during weekends disengage for different reasons than those who work during office hours. While the first group reported social reasons as their primary motivation to leave, the latter mostly cited job-related reasons. In our previous work~\citep{soheal2019}, we uncovered potential reasons behind developers' inactive periods, including personal (e.g., job change, financial issues, lack of interests) and project-related (e.g., role change, governance issues) reasons. \revisionone{Our current study further investigates these reasons, quantifying how frequently they occur.  

Finally, \cite{decan2020gap} predict developers' imminent abandonment by considering their commit activity. Although we are not predicting abandonment, our results can inform similar prediction models by bringing to light the importance of the work rhythm and by characterizing multiple states and their transitions, which consider coding and non-coding activities.} 

In summary, existing research has focused on studying the health of OSS communities concerning developers' onboarding, retention, and turnover. While a few papers consider the reasons behind OSS developers' disengagement, turnover, and the consequences of such actions, it is still unknown how breaks comprise part of the developer life cycle. \revisionone{The current literature does not explore the extent of disengagement, nor how to identify breaks based on the contribution rhythm of developers. In this paper, we broaden the literature by going beyond understanding the reasons to stay or leave. We investigate developers' disengagement, modeling intermediate states in which developers pause their coding contribution or engage in non-coding activities, and then either resume coding, or extend their hiatus until eventually abandoning a project altogether.} 

\revisionone{
\section{Research Framework}
\label{sec:res-framework}

The goal of this study is to investigate how core developers take breaks and disengage from OSS projects. To conduct the study, we devised a research framework divided into two phases (see Fig.~\ref{fig:res-workflow}).

\begin{figure}[hbt]
\centerline{
\includegraphics[width=1.1\textwidth]{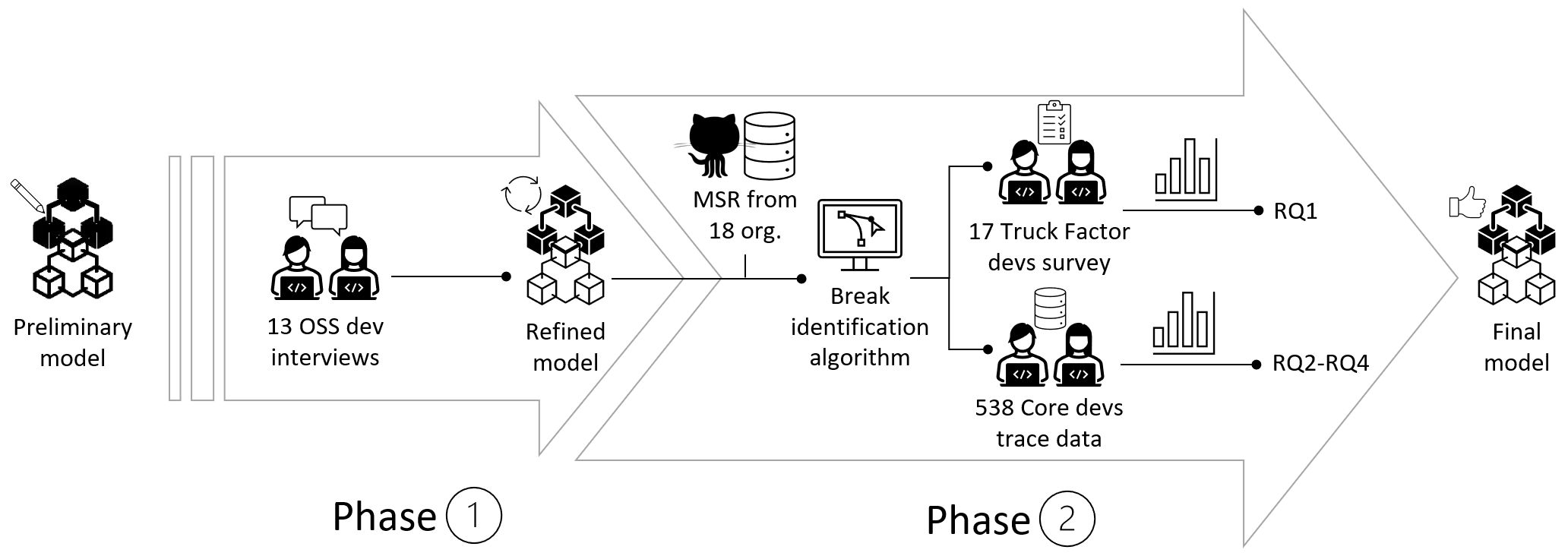}}
\caption{An overview of the research framework.}
\label{fig:res-workflow}
\end{figure}

In Phase~I (see Sect.~\ref{sec:phase_one}), we refined the preliminary model of developers' activities and breaks from our previous work \citep{soheal2019}. Accordingly, we carried out interviews with 13 OSS developers from 8 projects. Building on their feedback and insights, we designed a revised state model. 

In Phase~II (see Sect.~\ref{sec:phase_two}), we collected a larger sample of 18 projects from \textsc{GitHub}. Building on the revised model, we designed and implemented an algorithm that analyzes trace data mined from project repositories to establish the working rhythm of OSS developers and identify their states (activities and breaks) and transitions, at the organization level, as defined in the model.

To further understand the phenomenon of core developers taking breaks from and abandoning OSS projects, we defined the following research questions:
\\\noindent\textit{\textbf{RQ1.} To what extent can our method identify developers' breaks?}
\\\noindent\textit{\textbf{RQ2.} How often do core developers take breaks?}
\\\noindent\textit{\textbf{RQ3.} How long do core developers' breaks last?}
\\\noindent\textit{\textbf{RQ4.} How common are the transitions between states?}

To answer RQ1, we applied the Truck Factor algorithm~\citep{avelino2016novel} to identify 75 core developers of the selected projects. Then, we invited those with a public email address to participate in a personalized survey where we asked them to provide feedback on the revised model and to acknowledge their own breaks and transition as identified by our algorithm. We received answers from 17 developers. The results of this step were used to further refine the model.

For the remaining research questions RQ2-4, we considered a larger sample of core developers, i.e., those who authored 80\% of a project codebase, because, as discussed in Sect.~\ref{sec:core-devs}, the Truck Factor algorithm is far more restrictive. We ran our break-identification algorithm on the trace data of 538 core developers to quantitatively assess the phenomenon of inactivity and disengagement in OSS projects on a larger scale.

Next, we detail the activities carried out in the two phases of the research framework.
}

\section{Phase I - Modeling Developers' Activities and Breaks}
\label{sec:phase_one}

\begin{figure}[tb]
\centerline{
\includegraphics[width=.65\textwidth]{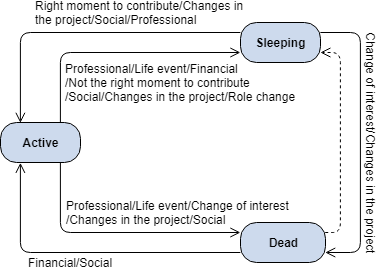}}
\caption{The original model of OSS developers' states and contribution breaks \citep{soheal2019}.}
\label{fig:original-statemodel}
\end{figure}

In our previous work~\citep{soheal2019}, we designed a preliminary state model, which defines states for project disengagement and explains the reasons for the transitions, as can be seen in Fig.~\ref{fig:original-statemodel}. We designed this model by conducting six interviews with experienced OSS developers and comparing their commit timeline on \textsc{GitHub} to the circadian rhythm (or sleep-wake cycle). \revisionone{The interviewed developers contributed to different projects, including Radar Parlamentar,\footnote{\url{https://gitlab.com/radar-parlamentar/radar}} Noosfero,\footnote{\url{https://gitlab.com/noosfero/noosfero}} Crossminer,\footnote{\url{https://www.crossminer.org}} GrimoireLab,\footnote{\url{https://chaoss.github.io/grimoirelab}} KDE Cantor,\footnote{\url{https://edu.kde.org/cantor}} and R.\footnote{\url{https://www.r-project.org}}}

In this preliminary model, nodes represent the developers' states in a project and edges represent the transitions between these states. The model is framed within the the metaphor of the circadian rhythm: the \textit{wake} stage---where intense brain activity is registered---corresponds to the \texttt{active} state, in which developers actively contribute source code changes; the \textit{sleep} stage---when brain activity is low, but other life signals are still observable---corresponds to the \texttt{sleeping} state, when developers pause their code contributions but still give signals of their presence in the community (e.g., by commenting issues and pull requests); finally, the \texttt{dead} stage corresponds to the state where a contributor has completely abandoned a project (i.e., neither code contributions nor signals of non-coding activities are found). The analysis of the interviews also helped to uncover a list of reasons for transitioning from one state to another. For example, a \texttt{sleeping} developer may wake up and return to \texttt{active} by contributing code after a break due to either a personal (e.g., life event, financial, lack of interests) or project-related (e.g., role change, social, governance issues) reason.

Because this model was based on a small set of interviews, we decided to further assess its validity and refine it before using it in the second phase of this study. 

\subsection{Evaluation of the Preliminary Model}
\label{sec:eval-preliminary-model}

To evaluate our preliminary model, we presented our results to 13 OSS developers from 8 OSS projects hosted on \textsc{GitHub}. These developers are long-term contributors and active in these projects. They were recruited from our personal network and referrals. \revisionone{These developers are from a different set of projects compared to those interviewed to conceive the preliminary model. The projects they contribute to (i.e., rails/rails,\revisiontwo{\footnote{\url{https://github.com/rails/rails}}} laravel/framework,\revisiontwo{\footnote{\url{https://github.com/laravel/framework}}} elixir-lang/elixir,\revisiontwo{\footnote{\url{https://github.com/elixir-lang/elixir}}} JabRef/jabref,\revisiontwo{\footnote{\url{https://github.com/JabRef/jabref}}} github/linguist,\revisiontwo{\footnote{\url{https://github.com/github/linguist}}} atom/atom,\revisiontwo{\footnote{\url{https://github.com/atom/atom}}} flutter/flutter,\revisiontwo{\footnote{\url{https://github.com/flutter/flutter}}} and ionic-team/ionic-framework)\revisiontwo{\footnote{\url{https://github.com/ionic-team/ionic-framework}}} are active, popular, with 5+ years of historic records, and diverse in terms of programming languages. Table~\ref{tab:aboutProjects} includes the details of these projects (which are marked with a \textsuperscript{*} in the table), among others.} 

We tailored a structured interview (questionnaire) for each respondent. We first collected the project history of each respondent from \textsc{GitHub}---using the \textsc{GitHub} API---and instantiated the model for that data. To identify breaks, we applied a method based on the \textit{far-out values} approach used for computing outliers (detailed in Sect.~\ref{sec:fov}), looking for breaks outside the developers' natural contribution rhythm. We also asked the respondents to give feedback about the state transitions (we showed up to 3 examples of each transition) and about the model as a whole. 

After analyzing their answers, we identified three recurring issues. The first issue regarded considering the states based on a single project instead of analyzing the entire organization, as some developers appeared as \texttt{sleeping}, but in fact are contributing to other repositories within the same organization. The second issue was that the term \texttt{sleeping} was misleading, as it failed to capture that, albeit not coding, many members are still actively contributing, e.g., by managing issues or the project. The third problem was also related to naming, specifically of the \texttt{dead} state; some respondents pointed out that the term was too negative and `terminal,' therefore failing to capture that some disengaged developers can actually go back. As one of the interviewees commented, a \texttt{dead} developer would be one who ``\textit{does not want to contribute anymore to that community ever again}.'' Instead, the developers also noticed a high number of transitions between \texttt{dead} and \texttt{active} states as well as between \texttt{active} and \texttt{sleeping}. 

Based on the feedback, we defined a revised version of the model.

\subsection{The Revised Model of Inactivity}
\label{sec:revised_model}

\begin{figure}[t]
\centerline{
\includegraphics[width=.7\textwidth]{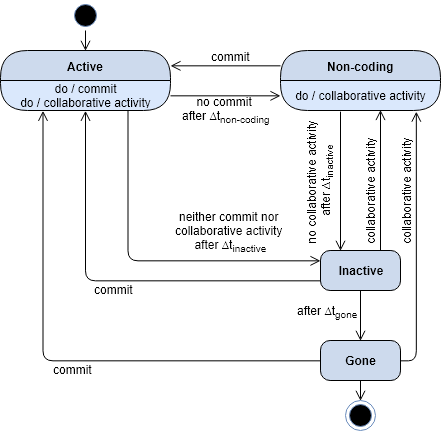}}
\caption{The state diagram of the revised model. A transition label represents the event triggering the state change, whereas internal activities are performed while in state.}
\label{fig:statemodel}
\end{figure}

Given the feedback from the structured interviews, we defined a revised model (see Fig.~\ref{fig:statemodel}) to inform this study. Compared to the preliminary version (Fig.~\ref{fig:original-statemodel}), we dropped the metaphor of the circadian rhythm; as such, we renamed the \texttt{dead} state as \texttt{gone}, and \texttt{sleeping} as \texttt{non-coding}. \revisiontwo{We explicitly added this state because contributors follow different pathways to contribute and contributors not working with code are usually `hidden'~\citep{trinkenreich2020hidden}.} In addition, we added a new state between \texttt{non-coding} and \texttt{gone}, named \texttt{inactive}. 

Also, the revised model includes transition labels that indicate the \ul{event} that triggers the state change; therefore, if active developers do not perform any \ul{commit} in a given time interval (i.e., \ul{no commit after $\Delta t_{non-coding}$}), but perform other activities, they transition from \texttt{active} into the \texttt{non-coding} state. The classification of \texttt{coding} and \texttt{non-coding} activities is provided later in Sect.~\ref{sec:labeling}, along with the description of the algorithm for labeling breaks and transitions accordingly.

OSS developers who take breaks from writing code may resume their \ul{commit} activity at some point. However, if \texttt{non-coding} developers do not resume code contribution but rather stop contributing (e.g., no issue comments, no pull requests reviews) for a time interval (i.e., \ul{no collaboration activity after $\Delta t_{inactive}$}), they become \texttt{inactive}. Entering into such a state is also possible for \texttt{active} developers, should they stop both committing and contributing by other means altogether (i.e., \ul{neither commit nor collaborate via issue tracker for at least $\Delta t_{inactive}$}). Developers can come out of the \texttt{inactive} state by resuming their activities. Otherwise, if their hiatus continues for at least $\Delta t_{gone}$, they eventually will be considered \texttt{gone} (i.e., they abandoned the project). As such, the transition towards \texttt{gone} is only possible from \texttt{inactive}. Finally, developers can go back to either \texttt{active} (by resuming code commits) or \texttt{non-coding} state (by reviewing code, opening issues, commenting, etc.).

\section{Phase II - Evaluation of the Revised Model and Characterization of the Phenomenon}
\label{sec:phase_two}

In this section, we describe the method adopted for the second phase of our research, including the data collection and sampling, the definition of break periods, and the identification of transitions between states. 

\subsection{Data Collection and Sampling}\label{sec:data-collection}

We mined project data from eighteen projects from different organizations hosted on \textsc{GitHub}. We extracted their entire history of activities, including commits, issues, and pull requests. The dataset and all the scripts for data collection and analysis are available online as supplementary material.\footnote{\url{https://doi.org/10.5281/zenodo.3731344}}  

\revisionone{
We used convenience sampling to select the organizations (see Table~\ref{tab:aboutOrganizations}) and projects (see Table~\ref{tab:aboutProjects}) used in the study. We started by including the same eight projects and organizations used for the preliminary model evaluation (see Sect.~\ref{sec:eval-preliminary-model}), namely rails/rails, laravel/framework, elixir-lang/elixir, JabRef/jabref, github/linguist, atom/atom, flutter/flutter, and ionic-team/ionic-framework (which are identified with a * in Table~\ref{tab:aboutProjects}). This choice was motivated by the fact that we have connections in those organizations, which we were planning to use to recruit developers for future qualitative assessments.

To select the other ten projects and their organizations, from the `Topics' section on the \textsc{GitHub} website,\footnote{\url{https://github.com/topics}} we identified the ten most trending topics and then moved on to select up to three projects per topic, excluding \textit{a priori} research projects and small, one-person projects. Instead, we opted to select active, successful projects whose discontinuation would pose a threat to the sustainability of many others depending on it.
We also filtered out projects within organizations with too many other projects (>350) to contain the extraction time within acceptable limits; likewise, we filtered out projects deemed too large in terms of the number of distinct contributors (>4000) or development history (>15 years).
Finally, we selected the highest-rated projects (\textsc{GitHub} stars) that would also add variety to the sample in terms of size (using the number of contributors, pull requests, and LOC as proxy), history (age), and programming language.

After completing the sampling, we ensured we selected the \textit{main} project from each organization. The identification was straightforward in most of the cases, as the largest project (in terms of the number of contributors, stars, forks received, pull requests, and LOC) is either homonymous with its organization (e.g., rails/rails, atom/atom) or has a closely resembling name (e.g., elixir-lang/elixir, ionic-team/ionic-framework). The verification in the case of github and facebook organizations, from which we respectively selected linguist and react, required extra care. We verified that the two projects were indeed the largest ones after excluding those that are mirrors or forks of other projects, as well as documentation-only projects (i.e., do not contain any source code).}

\begin{table}[t]
\caption{The eighteen \textsc{GitHub} organizations sampled for this study, sorted by number of repositories, and the selected projects (as of January 2020).} 
\label{tab:aboutOrganizations}
\centering
\footnotesize
\begin{tabular}{lrl}
\hline
\textbf{Organization} & \textbf{\#repos} & \textbf{Selected project} \\ \hline 
github          & 349   & linguist         \\ 
atom            & 256   & atom              \\ 
ionic-team      & 257   & ionic-framework   \\ 
nodejs          & 172   & node              \\ 
facebook        & 162   & react             \\ 
rails           & 98    & rails             \\ 
aseprite        & 64    & aseprite          \\ 
jekyll          & 53    & jekyll            \\ 
laravel         & 56    & framework        \\ 
jquery          & 45    & jquery           \\ 
MinecraftForge  & 40    & MinecraftForge    \\ 
JabRef          & 41    & jabref            \\ 
SpaceVim        & 35    & SpaceVim          \\ 
flutter         & 34    & flutter           \\  
fastlane        & 32    & fastlane          \\ 
crystal-lang    & 25    & crystal           \\ 
BabylonJS       & 21    & Babylon.js        \\ 
elixir-lang     & 9     & elixir            \\ \hline 
\end{tabular}
\end{table}

\begin{table}[t]
\caption{A breakdown of the eighteen \textsc{GitHub} projects analyzed for each organization in the study (as of January 2020). Projects with a \textsuperscript{*} represent those that also had developers interviewed to evaluate the preliminary model (in Phase 1).} 
\label{tab:aboutProjects}
\centering
\footnotesize
\begin{tabular}{lllllll}
\hline
\textbf{Project} & \textbf{\begin{tabular}[c]{@{}l@{}}Progr.\\ language\end{tabular}} & \textbf{Age} & \textbf{\begin{tabular}[c]{@{}l@{}}\#contri-\\ butors\end{tabular}} & \textbf{LOC} & \textbf{\begin{tabular}[c]{@{}l@{}}\#pull\\ reqs\end{tabular}} & \textbf{\#stars} \\ \hline
linguist\textsuperscript{*} &  Ruby; C              & 8 & 760   & 191k  & 2.7k  & 6.8k      \\ 
atom\textsuperscript{*}  & JavaScript            & 8 & 430   & 196k  & 4.4k  & 48.9k     \\ 
ionic-framework\textsuperscript{*} & TypeScript            & 4 & 328   & 115k  & 2.9k  & 38.2k     \\ 
node                     & JavaScript; C++       & 10& 2,471 & 5.9M  & 17.8k & 61.9k     \\ 
react                    & JavaScript            & 6 & 1,295 & 191k  & 8k    & 130.8k    \\ 
rails\textsuperscript{*} & Ruby                  & 15& 3,825 & 328k  & 23.5k & 43.3k     \\ 
aseprite                 & C++                   & 12& 40    & 227k  & 185   & 7.4k      \\ 
jekyll                   & Ruby                  & 11& 874   & 41k   & 3.6k  & 39.2k     \\ 
laravel/framework\textsuperscript{*}& PHP                   & 6 & 1,962 & 115k  & 16.4k & 17.5k     \\ 
jquery                   & JavaScript            & 13& 278   & 38k   & 2.5k  & 52.7k     \\ 
MinecraftForge           & Java                  & 8 & 320   & 82k   & 3.2k  & 3.5k      \\ 
jabref\textsuperscript{*} & Java                  & 15& 213   & 138k  & 2.8k  & 1.3k      \\ 
SpaceVim                 & Vim                   & 3 & 186   & 221k  & 1.6k  & 12.6k     \\ 
flutter\textsuperscript{*} & Dart                  & 5 & 399   & 487k  & 12.9k & 66.8k     \\
fastlane                 & Ruby                  & 5 & 1,016 & 561k  & 5.6k  & 27.5k     \\ 
crystal                  & Crystal               & 7 & 347   & 15k   & 3.5k  & 14.1k     \\ 
Babylon.js               & JavaScript; TypeScript& 6 & 256   & 2.6M  & 5.5k  & 1.5k      \\ 
elixir\textsuperscript{*}& Elixir                & 8 & 849   & 47k   & 5.3k  & 15.4k     \\ \hline

\end{tabular}
\end{table}


\revisionone{Looking at the breakdown of the sample, we observe that the selected organizations (Table~\ref{tab:aboutOrganizations}) host a number of repositories ranging between 9 and 349; regarding the projects (Table~\ref{tab:aboutProjects}), we notice that, as intended, they are written in different programming languages (e.g., Java, Ruby, PHP, Dart) and have a history (3-15 years) long enough to observe the development rhythm of its contributors. The projects vary also in terms of size (40-3.8k contributors, 1.5k-5.9M LOC, 1.6k-23.5k pull requests) and popularity (1.3k-131k \textsc{GitHub} stars). 
}

In the rest of the paper, we focus on studying the core developers of the selected projects while analyzing their (in)activity across all projects within the entire organization. \revisiontwo{This is motivated by the observation that---albeit most of their commit activity focuses on the main project---on average nearly 36\% of the total commits by the sampled core developers are contributed to the other projects from the same organization}.

\subsubsection{Identifying Core Developers and Their Pauses}
\label{sec:core-devs}

For each project in the selected organizations, we first used the \textsc{GitHub} API to collect the commit history of all developers. Then, for each developer, we checked the days when they made a commit to the projects and called them `commit days.' Accordingly, we define a `pause' as the interval (in days) between two consecutive commit days. For each project, we collected the number of pauses taken by all the developers who contributed at least twice (i.e., we observed those who took at least one pause) and normalized it by the number of years that they have been in the project.

\begin{table}[t]
\caption{\textit{Truck Factor developer}s (TF) and \textit{Core developers responsible for the 80\% of total commits} (Core), extracted for each of the projects in the sample.}
\begin{center}
\footnotesize
\label{tab:tfcomparison}
\begin{tabular}{lrrrc}
\hline
\textbf{Project}              & \textbf{Devs} & \textbf{TF} & \textbf{Core} & \textbf{\% TF in Core} \\ \hline
nodejs/node                & 2,031 & 15    & 115   & 87\%  \\
rails/rails               & 3,834 & 13    & 93    & 100\% \\
laravel/framework   & 1,979 & 1     & 74    & 100\% \\
facebook/react               & 1,301 & 4     & 33    & 100\% \\
fastlane/fastlane            & 958   & 3     & 19    & 100\% \\
elixir-lang/elixir              & 851   & 1     & 13    & 100\% \\ 
jekyll/jekyll              & 846   & 2     & 17    & 100\% \\
github/linguist            & 765   & 3     & 89    & 100\% \\ 
atom/atom                & 431   & 4     & 11    & 25\%  \\
flutter/flutter             & 402   & 7     & 28    & 100\% \\ 
ionic-team/ionic-framework  & 331   & 3     & 8     & 100\% \\
crystal-lang/crystal             & 327   & 2     & 6     & 50\% \\
MinecraftForge/MinecraftForge      & 289   & 3     & 6     & 100\% \\
jquery/jquery              & 263   & 3     & 9     & 100\% \\
BabylonJS/Babylon.js          & 231   & 2     & 7     & 100\% \\
JabRef/jabref              & 204   & 7     & 8     & 71\%  \\
SpaceVim/SpaceVim            & 158   & 1     & 1     & 100\% \\
aseprite/aseprite            & 37    & 1     & 1     & 100\% \\\hline
            & \textit{Tot.} & 75    & 538   &        \\ \hline
\end{tabular}
\end{center}
\end{table}

The focus of our analysis is on core contributors because their disengagement poses a serious risk to the survival of OSS projects~\citep{avelino2019}. Therefore, we applied the Truck Factor (TF) algorithm~\citep{avelino2016novel} to select the subset of the developers who are key to the project, and whose disengagement would greatly affect project survival (see Table~\ref{tab:tfcomparison}). As the Truck Factor algorithm is not the only approach for identifying key people from the projects, we also used a popular heuristic based on the number of commits from the project contributors, which usually follows a heavy-tailed distribution \citep{coelho2018}. Thus, in Table~\ref{tab:tfcomparison}, we also report the core developers of a project, those who produced 80\% of the total commits in a project (Commit-Based Heuristic). We noticed that the Truck Factor algorithm indeed is more selective than using the Commit-Based Heuristic and that most of the time the Truck Factor developers of a project are also included in the set of core contributors identified using the Commit-Based Heuristic. The only exceptions (i.e., nodejs/node, JabRef/jabref, and atom/atom) are caused by the fact that, unlike the Commit-Based Heuristic, the implementation of the Truck Factor approach that we used does not take into account the commits made on documentation files.

Because Truck Factor is more selective, we used this approach to identify the developers of each main project to assess our model (RQ1). Thus, we collected their data and analyzed their inactivity according to the state model. We assessed the outcomes of the model by conducting a survey with the Truck Factor developers, aiming to validate our findings.

After assessing the model, we used the Commit-Based Heuristic approach to broaden the understanding of the phenomenon. With the complete set of core developers identified, we conducted a quantitative analysis based on the data collected to answer RQ2, RQ3, and RQ4.

\subsubsection{Identifying Core Developers' Inactivity Periods}
\label{sec:fov}

Because each developer has their own rhythm of development, not every pause in between two consecutive commits indicates that a developer has become inactive. For instance, for a developer who typically commits every other day, a week with no contribution to any project of an organization represents an actual period of inactivity; by comparison, a week without a contribution may not mean inactivity for a developer whose rhythm comprises only a few commits per month. Therefore, to identify the actual \textit{inactivity periods} for each developer, we first create an individual array with all their pauses (in days) at the organization-level (i.e., for all the projects therein). Since developers' pauses vary in length, we only consider as inactivity periods those that are `longer than usual.'

\begin{figure}[t]
\centerline{
	\includegraphics[width=.72\textwidth]{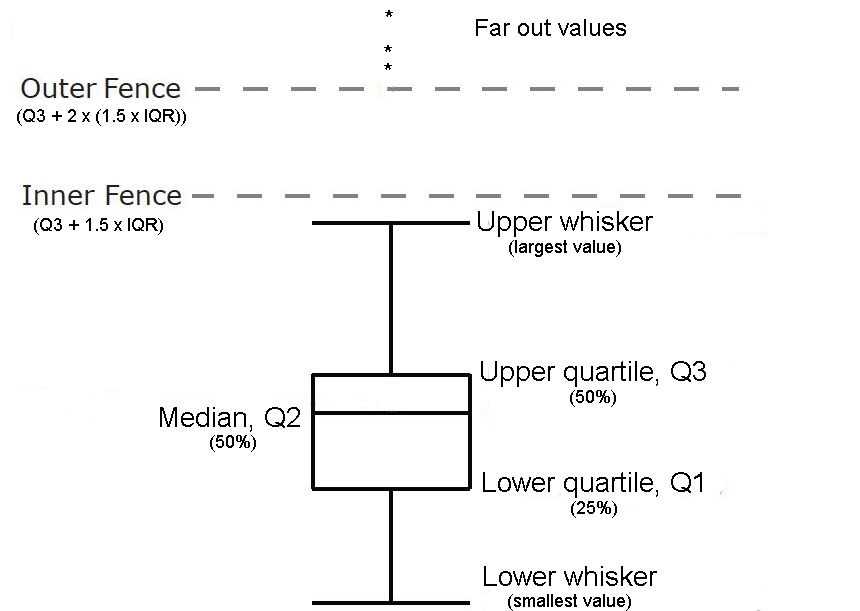}}
\caption{An example of box plot with far out values above the outer fence.}
\label{fig:faroutvalue}
\end{figure}

To identify longer-than-usual pauses, we set a developer-specific threshold $T_{fov}$ using the \textit{far out values} approach for detecting outliers in a distribution~\citep{book:Tukey1977}. A \textit{far out value} is defined as an extreme value larger than the upper quartile plus 3 times the interquartile range, i.e., the outer fences in a box plot (see Fig.~\ref{fig:faroutvalue}). In our case, let $P=<p_1,p_2,...,p_n>$ be the array of all the pauses taken by a given developer, $Q_1(P)$ the first quartile of the distribution, $Q_3(P)$ the third quartile, and $IQR=Q_3(P)-Q_1(P)$ the interquartile range (or \textit{H-spread}). Thus, based on the previous definition, we calculate the developer-specific threshold as $T_{fov}=Q_3(P)+3 \times IQR$.

The rhythm of a developer is not only unique, but also likely to change over time. Furthermore, the lifespan of participation in a project and its organization also varies from developer to developer. Therefore, it was necessary to identify a time-interval long enough to observe the natural rhythm of contributions and inactivity periods, and the `variations' from it. As such, to account for these variations, we refined our approach based on the far out values by introducing a \textit{sliding window} mechanism. 

\begin{figure}[t]
\centerline{
	\includegraphics[width=.85\textwidth]{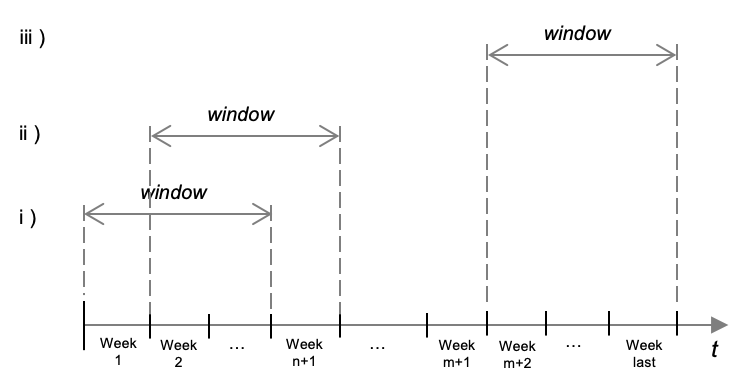}}
\caption{Inactivity periods are identified as the longer-than-usual pauses within each sliding window of size $w$ and a forward shift of 1 week. We tested windows sizes of $w=1,3,4,6,12$ months and selected 3 months as the optimal configuration.}
\label{fig:slidingwindow}
\end{figure}

To define the size (in months) of this window, we experimented with different values, namely $w = 1,3,4,6,12$. For each value, we started by considering the initial window $w$ (see step \textit{i} in Fig.~\ref{fig:slidingwindow}), identified all the pauses therein, and computed the \textit{far-out values} threshold $T_{fov}$; then, we identified the inactivity periods as the longer-than-usual pauses that are above $T_{fov}$ within the window. After that, we moved the window $w$ forward by 1 week (see step \textit{ii}) and repeated the shift until the last week of the contribution history (step \textit{iii}), thus eventually extracting all the inactivity periods from a developer's commit history. We selected a window size of 3 months because there are no cases of Truck Factor developers with a history of contributions smaller than the selected window size, as is the case for $w=4,6,12$ months (i.e., no data point is lost because of a too large window size); at the same time, this value reduces the number of pauses that are longer than the window size, as occurs in the case of $w=1$ month (i.e., window size too small).

For further details, please refer to the complete algorithm available in Appendix~\ref{appendix:algo}.

\subsubsection{Labeling Breaks and Transitions}
\label{sec:labeling}

\revisionone{As noted by the participants in the preliminary model evaluation (see Sect.~\ref{sec:eval-preliminary-model}), developers often collaborate on multiple projects in the same organization. Therefore, broadening the activity analysis of the selected projects' core developers at the organization level was necessary for the completeness of our model. Indeed, today's OSS communities are more and more often examples of complex ecosystems of interrelated projects; for example, one developer might be  \texttt{non-coding} or \texttt{inactive} in the front-end project, while they are actively \texttt{coding} to fix a bug in the back-end project from the same organization. }

\revisionone{ Regarding the type of activities, we label as \texttt{coding} activity both making a commit to a local repository and opening a pull request. For pull requests, we point out that we analyze and include in the activity timeline all the commits therein, whether the pull request was closed (as merged or not) or still open at the time of data extraction. 

Other activities such as those related to code review and project management (e.g., closing a pull request, assigning an issue) are classified as \texttt{non-coding};} we included comments in pull request discussions, issues opened, comments in issue discussions, and other actions performed on issues (e.g., subscription, assignment, closing, labeling) that could be gathered via \textsc{GitHub} API.
Given the state diagram presented in Fig.~\ref{fig:statemodel}, we label as \texttt{non-coding} those inactivity periods when at least one non-commit action has been performed, whereas we call \texttt{inactive} breaks those periods of inactivity in which no events were extracted. 

Regarding the transitions, first we mined the list of commits and other collaboration events (e.g., pull request and issue comments) from \textsc{GitHub} and labeled the \textit{back-to-codings} (i.e., the transitions from \texttt{non-coding} to \texttt{active}), \textit{reactivations} (from \texttt{inactive} to \texttt{active} or \texttt{non-coding}), and \textit{comebacks} (from \texttt{gone} to \texttt{active} or \texttt{non-coding}). 

To label the remaining transitions, we defined the time intervals $\Delta t_{non-coding}$, $\Delta t_{inactive}$, and $\Delta t_{gone}$ after which a state change can be triggered if there were no commits and/or other events. Regarding the transitions to \texttt{non-coding} and \texttt{inactive}, we used $\Delta t_{non-coding} = \Delta t_{inactive} = T_{fov}$, the same developer-specific, far out value-based threshold used to select the breaks. The rationale behind this choice is to ensure that the time intervals account for the differences in the individual rhythm of contribution of each core developer. Additionally, this choice allows us to identify cases of multiple breaks and state transitions within one inactivity period. 

\begin{figure}[t]
    \centering
    \subfloat[An inactivity period containing one break labeled as \texttt{non-coding}.\label{fig:sleeping}]{%
       \includegraphics[width=0.85\textwidth]{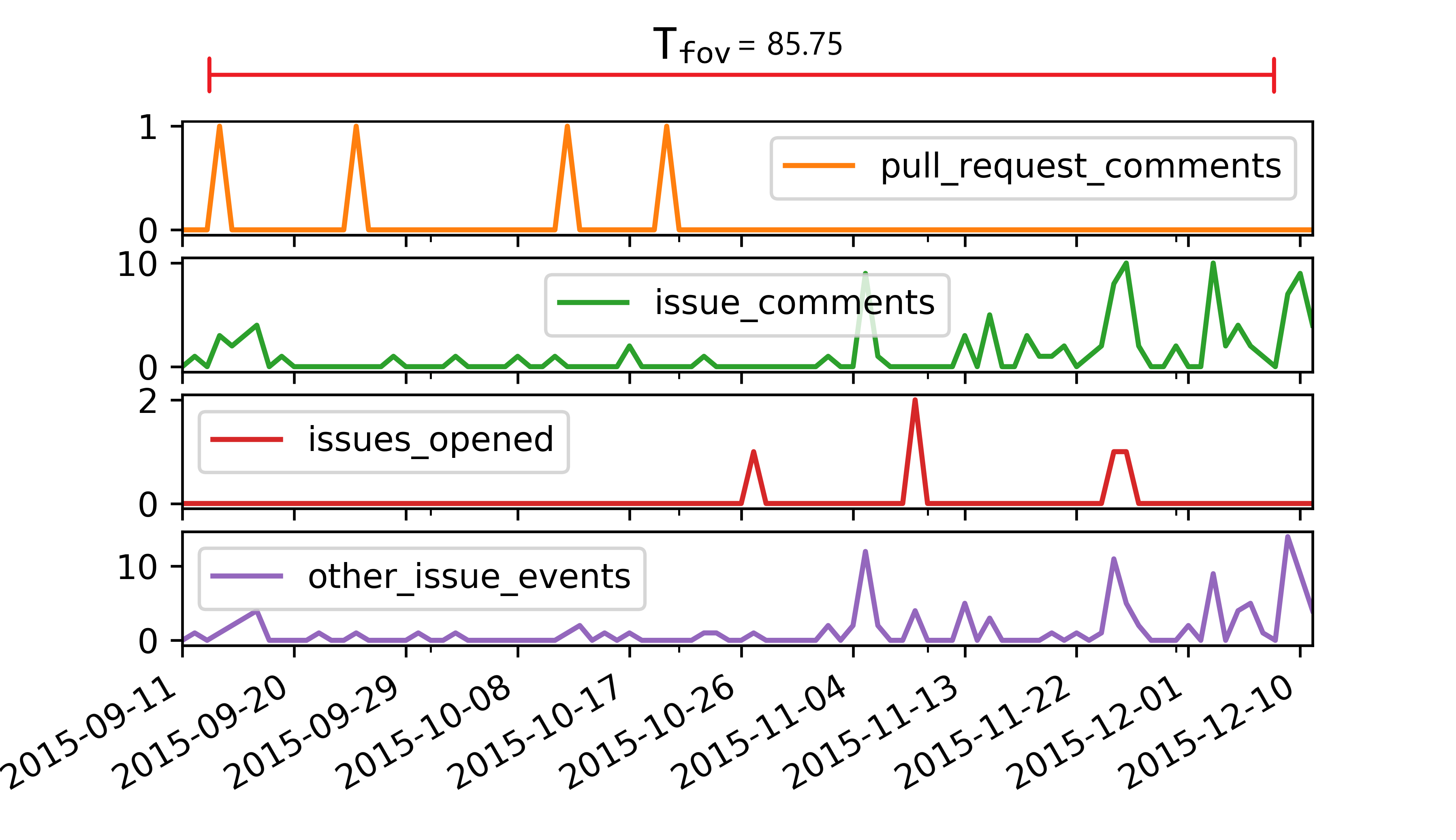}
     }
     \hfill
    \subfloat[An inactivity period containing a \texttt{non-coding} break that transitions into an \texttt{inactive} break.\label{fig:StoH}]{%
       \includegraphics[width=0.85\textwidth]{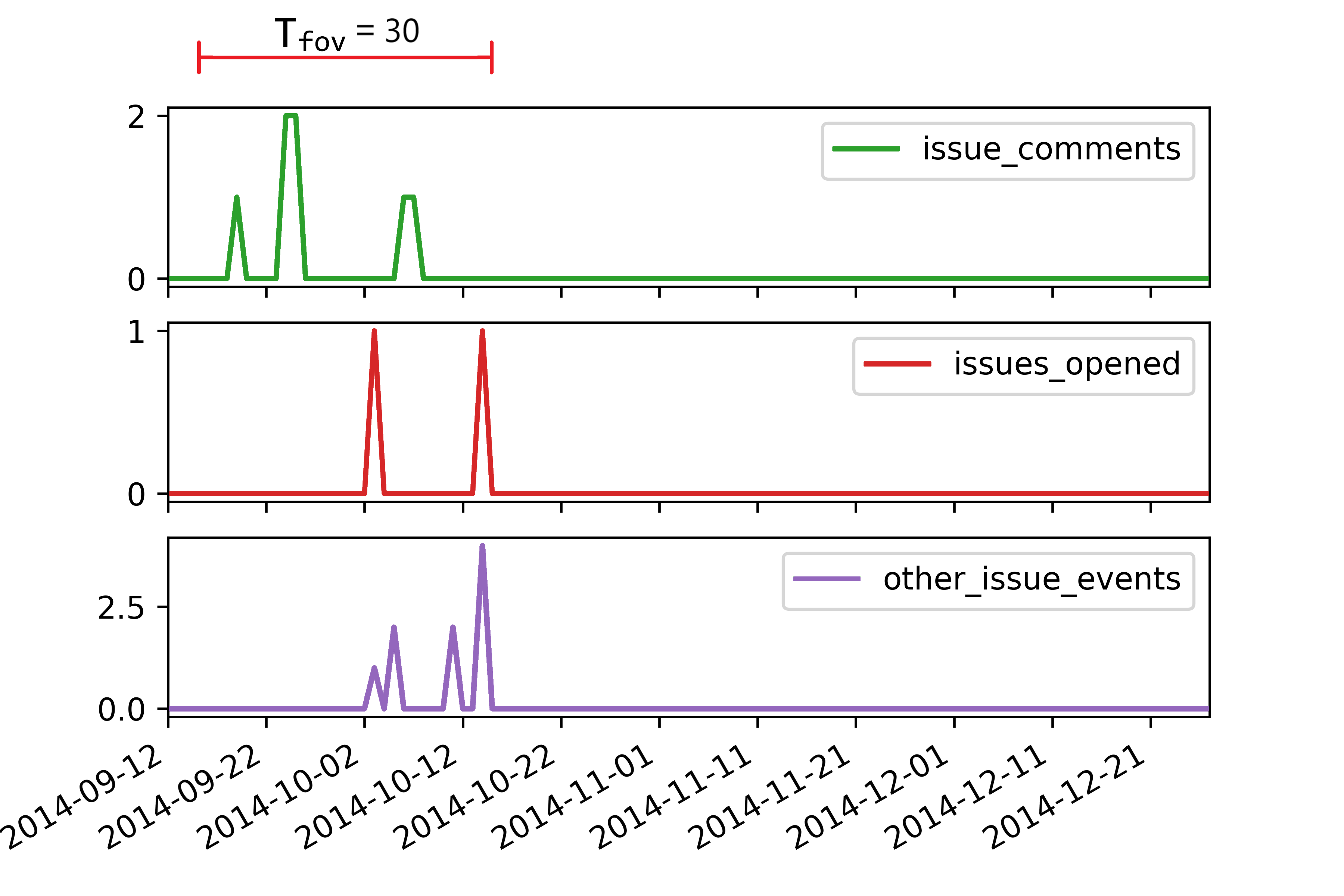}
     }
    \setlength{\belowcaptionskip}{-8pt} 
    \caption{Examples of how breaks are identified within inactive periods and labeled.}
    \label{fig:transitions_group}
\end{figure}

To further clarify these cases, we provide a couple of examples. Fig.~\ref{fig:sleeping} portrays an inactivity period within two consecutive commits made by a developer of the atom project on Sept. 11 and Dec. 13, 2015. Various collaboration traces are left by the developer during such an interval (e.g., pull requests and issues comments), each shown in the figure as separate graphs. The threshold $\Delta t_{non-coding} = \Delta t_{inactive} = T_{fov}$ is 85.75 days in this case. Because there are no sub-intervals longer than $T_{fov}$ during which any activity was performed, this inactivity period represents a single break, labeled as \texttt{non-coding}. Instead, looking at the inactivity period depicted in Fig.~\ref{fig:StoH}, we note that the developer was in the \texttt{non-coding} state for 32 days between Sep. 12 and Oct. 14, 2014. After that, (s)he did not leave further collaboration activity traces for longer than $T_{fov}$ = 30 days, thus transitioning to \texttt{inactive}. As such, this inactivity period contains two breaks: a \texttt{non-coding} break followed by an \texttt{inactive} break. 

Finally, regarding the extent of the hiatus $\Delta t_{gone}$ after which an \texttt{inactive} developer transitions into \texttt{gone}, existing studies rely on different thresholds to classify abandoning developers. As there is no consensus, we experimentally tested different thresholds. We first tested thresholds of three~\citep{constantinou2017} and six months~\citep{Miller2019}. However, they generated too many \texttt{gone} breaks and \textit{comeback} transitions. Then, we tested and selected $\Delta t_{gone}$ = 12 months as the threshold, which gave results in line with our model conceptualization, while also being consistent with the speculations reported in our previous work~\citep{soheal2019} that a developer's `death' occurs after a year of complete inactivity. \cite{avelino2019} also found that this threshold is the least sensitive to error when computing the Truck Factor of popular \textsc{GitHub} projects.

\section{Phase II - Results}
\label{sec:results}

\subsection{RQ1 -- Evaluation of the model and identification method}
\label{sec:res_RQ1}

To evaluate our model and identification method, we collected feedback from Truck Factor developers. Using the algorithm provided by \cite{avelino2016novel}, we identified 75 Truck Factor developers, distributed across the 18 projects, as shown in Table~\ref{tab:tfcomparison}. Among them, 53 have been \texttt{gone} and/or \texttt{inactive} at least once. Out of these, 34 had a valid email address publicly available on their \textsc{GitHub} page. 

Similarl to the preliminary evaluation (see Sect.~\ref{sec:eval-preliminary-model}), we designed personalized questionnaires\footnote{An example is available at \url{https://doi.org/10.6084/m9.figshare.12062964}} for each developer, asking them (1) to provide feedback about the model and 
(2) to confirm whether they recognize the breaks and transitions we identified. In particular, we selected up to nine breaks (three instances for each of the three types of break states) among the most recent ones to facilitate their recollection; then, we explained the model, as well as the meaning of each state, and for each break we asked: (i) \textit{Were you actually in that state?}, (ii) \textit{Why did you take that break?}, and (iii) \textit{Why did you return to commit/collaborate?}. We offered a \$15 gift card to those who completed the questionnaire. 

We received responses from 17 developers out of 34 invited (50\% of response rate). A breakdown of the respondents is reported in Table~\ref{tab:respondents_breakdown}. All respondents self-identified as male, with ages ranging between 21 and 50 (avg. 35). We received answers from developers of several projects: rails/rails (8), laravel/framework (2), flutter/flutter (2), JabRef/jabref (2), nodejs/node (1), elixir-lang/elixir (1), and github/linguist (1).


\begin{table}[tb]
\caption{Breakdown of survey respondents.}
\footnotesize
\centering
\begin{tabular}{llccc}
\hline 
\textbf{ID} & \textbf{Project} & \textbf{Age} & \textbf{Gender} & \textbf{\begin{tabular}[c]{@{}l@{}}\#  other projects in the \\ org. they contribute to\end{tabular}} \\ \hline
D-01         & nodejs/node                & 50        & Male              & -                               \\
D-02         & rails/rails                 & 30           & Male            & 1                                                       \\
D-03         & rails/rails                 & 32           & Male            & 1                                                       \\
D-04         & rails/rails                 & 47           & Male            & 0                                                       \\
D-05         & rails/rails                 & 45           & Male            & 2                                                       \\
D-06         & rails/rails                 & 23           & Male            & 7                                                       \\
D-07         & rails/rails                 & 32           & Male            & 1                                                       \\
D-08         & rails/rails                 & 32           & Male            & 2+                                                \\
D-09         & rails/rails                 & 37           & Male            & 0 
                                     \\
D-10         & laravel/framework               & 21           & Male            & 4                                                       \\
D-11         & laravel/framework               & 28           & Male            & 3                                                       \\
D-12         & JabRef/jabref                & 33        & Male              & 1                                                      \\
D-13         & JabRef/jabref                & 33        & Male              & 0                                                       \\
D-14         & elixir-lang/elixir                & 41           & Male            &  0                                              \\  
D-15         & facebook/linguist               & 40           & Male            & 0                                                       \\
D-16         & flutter/flutter               & 39           & Male            & all                              \\
D-17         & flutter/flutter               & 38           & Male            & 2+                                             \\

\hline
\end{tabular}
\label{tab:respondents_breakdown}
\end{table}

\noindent\textbf{Model feedback}. We asked the developers to rank their agreement with our inactivity model on a four-point Likert scale (1=\textit{Strongly disagree}, 2=\textit{Somewhat disagree}, 3=\textit{Somewhat agree}, and 4=\textit{Strongly agree}). We received 17 answers, of which only 1 \textit{strongly disagreed} (6\%); the remaining ranks are evenly distributed between \textit{moderately agree} (8, 47\%) and strongly agree (8, 47\%). The mean and median values are 3.4 and 3, respectively. The strong disagreement came from a developer who mentioned that ``\textit{\textsc{GitHub} doesn't track all the ways that I contribute to a project. For example, [...] I review design docs written in Google docs and I attend video conference meetings to mentor teammates}'' (D-16). This comment suggests that the name \texttt{non-coding} may unintentionally imply that developers are not actively participating unless they are coding. Consistently, albeit D-16 moderately agreed with model, D-13 commented that ``\textit{there's varying degree of `\texttt{active}.' Maybe this is more a continuum than one with discrete states}'', and D-10 suggested that we ``\textit{should also consider reviewing and filing bugs as \texttt{active}, in case you're not.}'' These observation are congruous with the comment from D-04, who observed that our model ``\textit{maybe makes the `commit' the ultimate contribution. For most projects that's the case, but [sometimes] the discussions around the new features added [...] are more important than the few line of codes that implement [them]. }''

We also received a few comments concerning the \texttt{gone} state. For instance, D-05 pointed that the name is maybe too strong and constrained: ``\textit{Per this definition, I am \texttt{gone} [...], but I'm still around, just no time to code or contribute. I occasionally do come to life, though}.'' In line with this, D-08 noted that ``\textit{you can be \texttt{inactive} for 12 months or longer without being \texttt{gone}. There's also a bit of an extra stage, which I guess I would just call `lurker,' where you're still watching everything happening, you are just not participating in it (unless it hits some personal threshold that pulls you back in to participate)}.'' This lurking stage matches our definition of \texttt{inactive}; as such, lurking may become an alternative name for \texttt{inactive} or be included in its the definition. Nevertheless, lurking represents a state from the perspective of the contributor, who can be viewed as \texttt{inactive} from the point of view of the project. 

\noindent\textbf{Breaks and state transitions acknowledgment}. We analyzed the questionnaire responses and, in particular, those cases where developers reported to disagree with the identified breaks and transitions. Table~\ref{tab:survey_breaks_trans_agreements} reports the breakdown of the acknowledgments with 88 breaks and 62 agreements overall (71\%). We observed 21 cases of disagreements (24\%) of which 10 (48\%) had ``\textit{don't know/can't remember}'' entered as a comment or simply had no comment. To uncover potential limitations of our model, next we focus on discussing the remaining cases of disagreement that were reported with a motivation.

Aligned with our previous observation, the \texttt{gone} breaks generated disagreements due to different perspectives. Two developers disagreed that they were \texttt{gone}. We analyzed their explanations and concluded that they were actually \texttt{gone} according to our definition, but not to their own. In fact, D-05 did not want to consider himself \texttt{gone} despite the fact that he did not contribute for over a year because he ``\textit{started several companies and work took over}.'' Similarly, D-04 reported that he simply ``\textit{did not find any bug or thing worthwhile contributing}.'' 

Regarding the \texttt{inactive} breaks, we found 8 disagreements out of 39 cases (20.5\%). In 6 cases, no comments were provided. In two cases, developers disagreed because they were studying new technologies (D-03) or doing design work (D-16) before contributing again. There are two cases we labeled as `other' (6.1\%): D-09 commented that he was sort of \texttt{inactive}: ``\textit{I stopped being employed to [work] on $<$project name$>$ full-time in 2010, but I remained pretty active as a mentor for at least one year after that}.'' As for the agreements (29, 74.4\%), the most cited reasons for \texttt{inactive} breaks are holidays, paid work, switching jobs, and personal matters.

Regarding \texttt{non-coding} breaks, we found 9 cases of disagreements out of 44 (20.5\%), 5 of which present no comments to analyze. D-05 disagreed that he was in the \texttt{non-coding} state. Unfortunately, his comment (``\textit{was working on other things''}) does not provide any clarification: if he was coding for any project of the organization, we would have found traces, because we mined the entire \textsc{GitHub} organization commit history; otherwise, any work on other, unrelated projects would not matter here. The remaining case concerned D-15, who explained his disagreement saying that he had ``\textit{already left the company}'' by then and, therefore, he could not possibly be contributing in any way to the project. The comment raised a red flag: having labeled it a \texttt{non-coding} break meant that we had found traces of collaboration activities through the \textsc{GitHub} API. After investigating, we found a couple of events that in which D-15 was mentioned in a PR conversion and assigned to an issue. As such, our speculation is that, unaware of his disengagement from the company, other project contributors kept mentioning him and assigning him to issues in his area of expertise. To avoid these cases, we fine-tuned our break labeling algorithm (see Appendix~\ref{appendix:algo}) to consider only those \textsc{GitHub} events that imply `active' participation of developers (e.g., writing a comment), discarding `passive' events such being mentioned in comment, being un/assigned from/to an issue. All the results reported next in this paper have been obtained after applying this change. Finally, as for the 3 remaining cases of disagreement labeled `other' (5.7\%), D-09 commented that he ``\textit{likely slowed down because the break was over the July 4$^{th}$ long weekend, but was still doing some work}.'' Yet, we could only retrieve traces of interaction on \textsc{GitHub} that do not include code contributions.

\begin{table}[tb]
\footnotesize
\centering
\caption{Break and state transition acknowledgments from the surveyed Truck Factor developers.}
\begin{tabular}{lrrrr}
\hline
\textbf{Breaks} & \textbf{Total} & \textbf{Agreements (\%)} & \textbf{Disagreements (\%)} & \textbf{Other} (\%) \\ \hline
Gone           & 5     & 1 (20\%)  & 4 (80\%)  & 0 (0\%) \\
Inactive       & 39    & 29 (74.4\%) & 7 (20.5\%)  & 2 (5.1\%) \\
Non-coding     & 44    & 32 (72.7\%) & 9 (20.5\%)  & 3 (6.8\%) \\\hline
\multicolumn{1}{r}{\textit{Tot.}}  & \textit{88}    & \textit{62 (70.5\%)} & \textit{21 (23.9\%)} & \textit{5 (5.7\%)} \\\hline 
\textbf{Transitions} & \textbf{Total} & \textbf{Agreements (\%)} & \textbf{Disagreements (\%)} & \textbf{Other} (\%) \\ \hline
Back-to-coding     & 40     & 33 (82.5\%)  & 5 (12.5\%)  & 2 (5\%) \\
Reactivation       & 40     & 30 (75\%)  & 8 (20\%)  & 2 (5\%) \\
Comeback           & 5      & 4  (80\%)  & 1 (20\%)  & 0 (0\%) \\\hline
\multicolumn{1}{r}{\textit{Tot.}}     & \textit{85}     & \textit{67 (78.8\%) } & \textit{14 (16.5\%)} & \textit{4 (4.7\%)}\\\hline
\end{tabular}
\label{tab:survey_breaks_trans_agreements}
\end{table}

Table~\ref{tab:survey_breaks_trans_agreements} also reports the breakdown of the 85 state transition acknowledgments, with 67 agreements overall (78.8\%). We observed 14 cases of disagreements, of which half had ``\textit{don't know/can't remember}'' entered as a comment or simply had no comment. Regarding \textit{back-to-coding} transitions, we found 5 cases of disagreement out of 40 (12.5\%), all of which with no comments. Regarding \textit{reactivations} (transitions from \texttt{inactive} or \texttt{non-coding} into \texttt{active}), we found 8 disagreements. D-03 pointed out that what we mined ``\textit{was probably a merge from an old pull request}.'' Yet, after checking, we did find a commit from him that he probably did not remember or missed while checking on \textsc{GitHub}. As for the 2 cases of partial disagreement with \textit{reactivation} transitions (i.e., classified as other), D-09 commented that, while he disagreed, he was not entirely sure. Again, we found commits authored by him thorough \textsf{git log}. Finally, with respect to the \textit{comebacks} from the \texttt{gone} state, we found 1 case of disagreement out of 5 (20\%), namely from D-05 who commented that he did not actually want to re-engage, but only ``\textit{needed to fix something in the code}.''

\noindent\textbf{Coding of the reasons why}. In our previous work \citep{soheal2019}, after interviewing OSS developers, we not only designed the preliminary version of the state model presented in Sect.~\ref{sec:phase_one}, but also defined a classification schema for the reasons why they took breaks. To gather a better understanding of the motivations of the developers surveyed in this study, we adapted the existing schema and developed two classifications to code, respectively, the comments to the the second question (i.e., ``\textit{Why did you take that break?},'' see Table~\ref{tab:codingbreaks}) and the third question (``\textit{Why did you return to $<$state$>$?},'' see Table~\ref{tab:codingtransitions}) of the questionnaire.

\begin{table}[t]
\caption{Coding schema for the reasons to take breaks.}
\label{tab:codingbreaks}
\begin{tabular}{lll}
\hline
\textbf{Category (\%)}        & \textbf{Reason (\%)}       & \textbf{Examples}                       \\ \hline
                             & Professional  (21.7\%)                                                          & \begin{tabular}[c]{@{}l@{}}Analysis/design work, job change, \\ learning a new technology, \\ student assignment\end{tabular} \\
                            \textbf{Personal-related}  & Life event  (14.5\%)                                                            & \begin{tabular}[c]{@{}l@{}}Vacation/holidays,\\ needed time to relax, child birth,\\ sickness, death in the family\end{tabular}                 \\
\textbf{(78.3\%)}             & Financial  (21.7\%)        & \begin{tabular}[c]{@{}l@{}}Need more time at work, \\ end of financial support \end{tabular}        \\
                           & Change of interest (20.3\%)                                                       & \begin{tabular}[c]{@{}l@{}}Grew uninterested, \\ migration to a new community \end{tabular}                                   \\ \hline
                            & \begin{tabular}[c]{@{}l@{}}Not the right time \\ to contribute (2.9\%)\end{tabular} & \begin{tabular}[c]{@{}l@{}} Did not feel responsible,\\ others took over\end{tabular}               \\
\textbf{Community-related} & Social (2.9\%)                                                                        & \begin{tabular}[c]{@{}l@{}}Problems with other members, \\ lack of feedback/recognition\end{tabular}                          \\
\textbf{(21.7\%)}                           & Changes in the project (2.9\%)                                                        & \begin{tabular}[c]{@{}l@{}}Technical, \\ organizational (governance)\end{tabular}                                             \\
                           & Role change (13\%)                                                                  & \begin{tabular}[c]{@{}l@{}}Becoming a project manager, \\ mentoring\end{tabular}                                              \\ \hline
\end{tabular}
\end{table}

\begin{table}[t]
\caption{Coding schema for the reasons to return to commit/contribute.}
\label{tab:codingtransitions}
\begin{tabular}{lll}
\hline
\textbf{Category (\%)}          & \textbf{Reason (\%)}      & \textbf{Examples}                                                                \\ \hline
 & Professional (50\%)          & \begin{tabular}[c]{@{}l@{}}Needed to fix something,\\ work duties, job change \end{tabular}  \\
\textbf{Personal-related} & Life event (4.3\%)             & End of vacation/holidays                                                         \\
\textbf{(82.5\%)}                            & Financial  (13\%)            & Doing paid work.                                                                   \\
                           & Interest in OSS  (15.2\%)     & \begin{tabular}[c]{@{}l@{}}Wanted/found opportunity\\ to contribute\end{tabular} \\ \hline
\textbf{Community-related} & Sense of responsibility  (13.1\%)        & \begin{tabular}[c]{@{}l@{}}Commitment,\\ exclusive knowledge\\feeling dependable\end{tabular}        \\
\textbf{(17.5\%)}   & Changes in the project (2.2\%) & Wanted to catch up  \\
& Role change (2.2\%) & Onboarding\\ \hline
\end{tabular}
\end{table}

In both coding schemas, we distinguished personal reasons (i.e., related to individual choices, needs, end events) from those concerning the community (i.e., events related to changes or interaction with others within the OSS community). We observe a consistent \tildex80/20\% split between the personal- vs. community-related categories.

Regarding the \textit{personal reasons for taking a break}, they are quite balanced, with \tildex{20\%} of the surveyed developers mentioning professional reasons (e.g., ``\textit{changed job}'', ``\textit{been busy doing analysis work}''), financial reasons (i.e., ``\textit{doing paid work}''), and a change of interest (e.g., ``\textit{started looking for other projects to contribute to}''). Slightly fewer developers (14.5\%) mentioned a life-related event as a cause for taking breaks (e.g., ``\textit{was on vacation/holidays}'', ``\textit{personal health-related issues}''), which are more varied. With respect to the \textit{community-related reasons for taking a break}, by far the most common is the change of role (13\%), such as becoming project manager or a mentor for newcomers. Changes in the organization of the community or personal issues with other members are not very common (\tildex{3\%}).

Regarding \textit{personal reasons for returning to commit or contribute}, by far the most common are professional (50\%), which are cited not only by paid developers---whose employer is sponsoring the project---but also by others who need to fix something in the code of the project that was blocking their work. The other cited reasons concern personal (15.2\%, e.g., ``\textit{found something to contribute to}'') and financial interest (13\%, e.g., ``\textit{was doing paid work}''). Again, life events such as end of vacation/holidays are the least common personal reasons (4.3\%) for returning among core developers. With respect to the \textit{community-related reasons for returning to commit or contribute}, the most common concern the sense of responsibility of the developer (\tildex{13\%}), who feels committed to the project and dependable, and realizes they hold exclusive, vital knowledge. The other reasons are related to changes of the developer's role or in the project itself, and are considerably less common (\tildex{2\%}).


\vspace{3mm}
\begin{tcolorbox}[standard jigsaw,
	title=RQ1 -- Evaluation of the model, opacityback=0]
	\small
	\begin{itemize}
		\item Almost all of the surveyed Truck Factor developers (94\%) agree with our state model of inactivity, either strongly (47\%) or moderately (47\%).
		\item Respectively, \tildex{71\%} and \tildex{79\%} of the surveyed developers acknowledge their breaks and state transitions. 
		\item The names chosen for the \texttt{active} and \texttt{non-coding} states can be misleading as they may unintentionally imply that developers are not actively participating unless they are coding.
		\item Most of the cited reasons among core developers for taking breaks as well as returning are personal (\tildex{80\%)} rather than related to the interaction within the community (\tildex{20\%}).
		\item The most common personal reasons for taking breaks are professional, financial, or lack of interest.
		\item The most common personal reason for returning to contribute are professional and the sense of responsibility toward the project.
	\end{itemize}
\end{tcolorbox}

\subsection{RQ2 -- Break Frequency}
\label{sec:res_RQ2}

To answer the research question about the frequency of inactivity periods in OSS development, we quantitatively analyzed the results of our approach to identify the \texttt{non-coding}, \texttt{inactive}, and \texttt{gone} breaks of the core developers from the eighteen sampled projects. To broaden our analysis, in the second phase of analysis we considered the set of 538 core developers identified using the Commit-Based Heuristic approach. Table~\ref{tab:inactive_cbh_developers} provides an overview of the core developers who have been inactive at least once, considering their activity in all the projects of each organization.

We notice that 97\% of the sampled core developers have been in the \texttt{non-coding} state at least once and 89\% have been \texttt{inactive}. Nearly a third (33\%) of the developers in our sample transitioned to the \texttt{gone} state at least once. On the one hand, we found projects in which no core developers have been in the \texttt{gone} state (i.e., aseprite, SpaceVim, and crystal-lang) or only one (i.e., ionic-team, BabylonJS, and MinecraftForge). On the other hand, in organizations like rails and github this happened for 65\% and 89\% of their core developers, respectively.

\begin{figure}[t]
\centerline{
\includegraphics[width=1.2\textwidth]{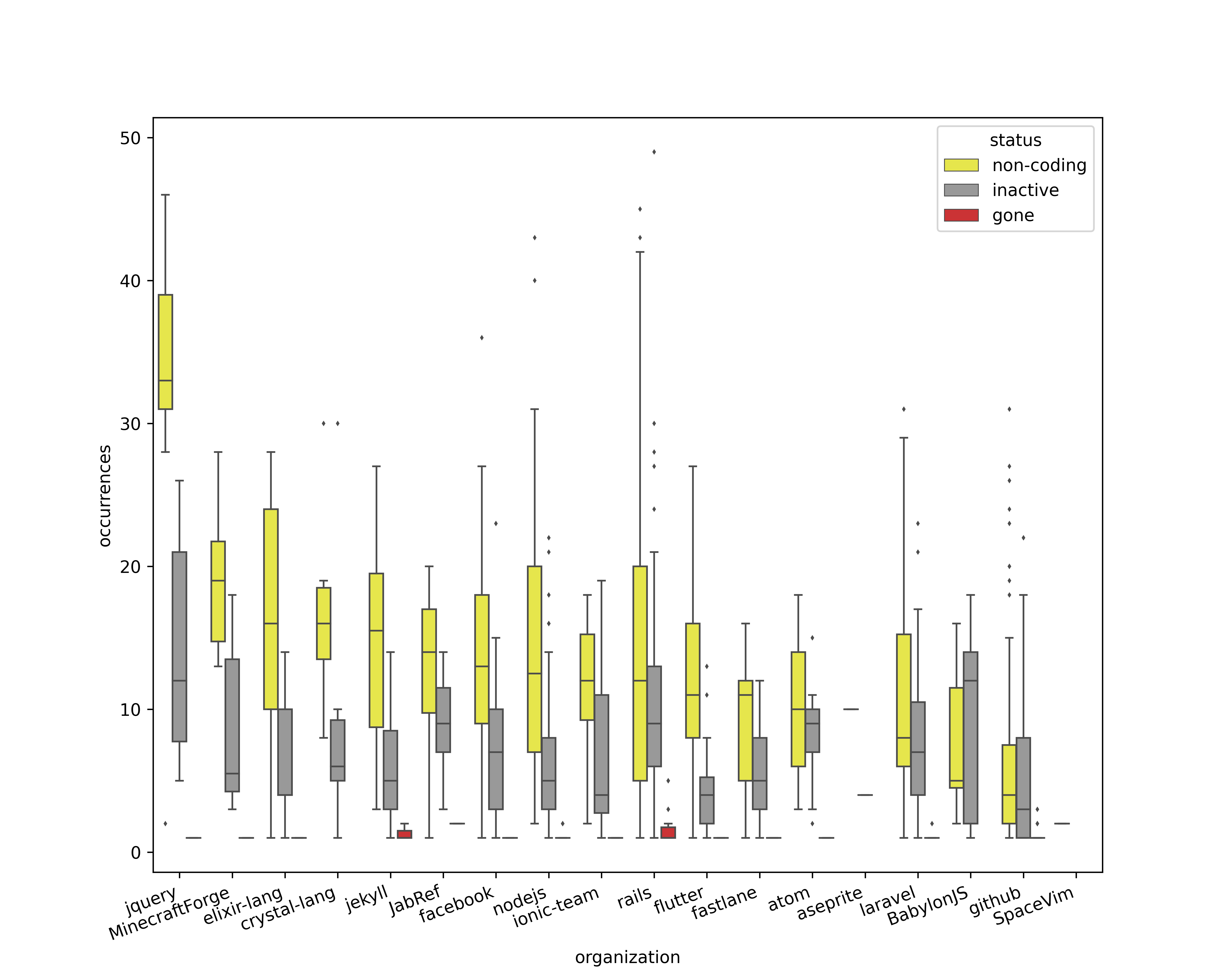}}
\setlength{\belowcaptionskip}{-17pt} 
\caption{Distributions of the average number of \texttt{non-coding}, \texttt{inactive}, and \texttt{gone} breaks per core developer in each organization. The boxplots are sorted in descending order with respect to the duration of non-coding breaks.}
\label{fig:inactivity_occurrences}
\end{figure}

Finally, in Fig.~\ref{fig:inactivity_occurrences}, we report the distributions of the average number of \texttt{non-coding}, \texttt{inactive}, and \texttt{gone} breaks per core developer in each organization. As expected, we observe that the \texttt{non-coding} (mean: 11.63, median: 10, SD: 9.14) periods occur more frequently than \texttt{inactive} breaks (mean: 6.56, median: 5, SD: 5.84), and that \texttt{gone} breaks are the least common (mean: 0.27, median: 0, SD: 0.60).

\vspace{3mm}
\begin{tcolorbox}[standard jigsaw,
	title=RQ2 -- Break frequency, opacityback=0]
	\small
	\begin{itemize}
		\item Almost all core developers take breaks, as 97\% and 89\% of them have been in the \texttt{non-coding} and \texttt{inactive} states at least once, respectively.
		\item About one third of the developers analyzed (33\%) have been \texttt{gone} at least once and disengaged completely from a project for one year or longer.
		\item \texttt{Non-coding} periods occur more frequently than \texttt{inactive} breaks within organizations; \texttt{gone} breaks are by far the least common.
	\end{itemize}
\end{tcolorbox}

\begin{table}[t]
\caption{Core Developers who have been inactive per organization.}
\footnotesize
\centering
\begin{tabular}{lrrr|c}
\hline
     & \multicolumn{3}{c|}{\textbf{\# developers who have been}}& \multicolumn{1}{c}{\textbf{\# developers}}\\
\textbf{Organization}     & \textbf{\texttt{Non-coding}}  & \textbf{\texttt{Inactive}}   & \textbf{\texttt{Gone}} & \textbf{\texttt{Gone} during data collection}      \\ \hline
nodejs          & 112 (97\%)    & 101 (88\%)     & 24 (21\%)   & 16 (14\%)\\
rails           & 92 (99\%)     & 93 (100\%)     & 60 (65\%) & 38 (41\%)\\
aseprite        & 1 (100\%)     & 1 (100\%)     & 0 (0\%)   & 0 (0\%)\\
jekyll          & 16 (94\%)    & 15 (88\%)     & 8 (47\%)  & 8 (47\%)\\
laravel         & 72 (97\%)     & 71 (96\%)     & 23 (31\%) & 11 (15\%)\\
MinecraftForge  & 6 (100\%)     & 6 (100\%)      & 1 (17\%)   & 0 (0\%)\\
JabRef          & 8 (100\%)     & 7 (88\%)      & 4 (50\%)  & 3 (38\%)\\
SpaceVim        & 1 (100\%)     & 0 (0\%)       & 0 (0\%)   & 0 (0\%)\\
fastlane        & 17 (89\%)    & 17 (89\%)     & 12 (63\%)  & 9 (47\%)\\
crystal-lang    & 6 (100\%)     & 6 (100\%)      & 0 (0\%)   & 0 (0\%)\\
BabylonJS       & 7 (100\%)     & 7 (100\%)     & 1 (14\%)   & 1 (14\%)\\
elixir-lang     & 13 (100\%)    & 13 (100\%)     & 5 (38\%)  & 2 (15\%)\\ 
github          & 68 (76\%)     & 64 (72\%)     & 79 (89\%) & 66 (74\%)\\
atom            & 10 (91\%)     & 11 (100\%)    & 6 (55\%)  & 5 (45\%)\\
ionic-team      & 8 (100\%)     & 8 (100\%)      & 1 (13\%)  & 1 (13\%)\\
facebook        & 33 (100\%)    & 33 (100\%)     & 10 (30\%)   & 7 (21\%)\\
jquery          & 9 (100\%)     & 8 (89\%)      & 4 (44\%)  & 3 (33\%)\\
flutter         & 27 (96\%)    & 25 (89\%)     & 5 (18\%)   & 3 (11\%)\\ \hline
\multicolumn{1}{r}{\textit{Avg.}}    & \textit{97\%} & \textit{89\%} & \textit{33\%} & - \\ \hline
\end{tabular}
\label{tab:inactive_cbh_developers}
\end{table}

\begin{figure}[t]
\centerline{
\includegraphics[width=1.2\textwidth]{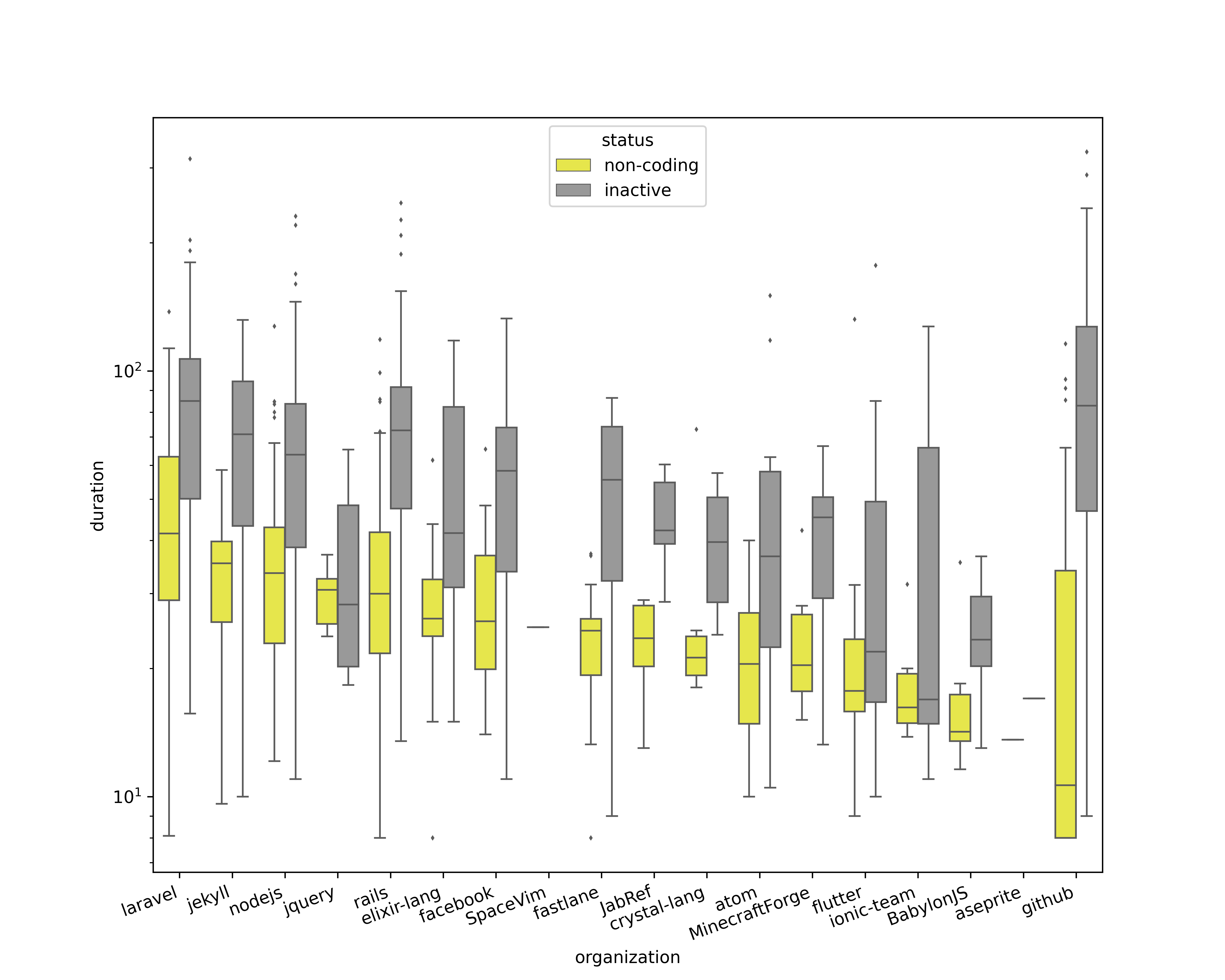}}
\caption{Distribution of the median duration (in days on a logarithmic scale) of \texttt{non-coding} vs. \texttt{inactive} breaks for each developer. The boxplots are sorted in descending order with respect to the  duration of non-coding breaks.}
\label{fig:sleep_hiber_durations}
\end{figure}

\subsection{RQ3 -- Break Length}
\label{sec:res_RQ3}

We analyzed the duration (in days) of the breaks, except for \texttt{gone} breaks, since most of them (see Table~\ref{tab:inactive_cbh_developers}) were still in progress at the time of this analysis. The breaks range from 367 to 2,893 days. Fig.~\ref{fig:sleep_hiber_durations} shows the boxplots of the average duration of the \texttt{non-coding} and the \texttt{inactive} breaks per core developer, grouped by organization and on a logarithmic scale.

Overall, the average duration of a \texttt{non-coding} break ranges from 8 to 994 days (mean: 32.26, median: 22, SD: 32.03), whereas the duration of an \texttt{inactive} break varies in the range of 8 to 374 days (mean: 65.59, median: 41, SD: 66.43). From Fig.~\ref{fig:sleep_hiber_durations}, we observe that the median duration of \texttt{inactive} breaks is consistently longer than that of \texttt{non-coding} breaks across all the organizations, with the sole exception of jquery. 

\revisionone{To understand whether these differences are statistically significant, for each organization we filtered those developers who have been in both states, and then performed a series of paired tests for each organization. Because the distribution of break durations are not normal, we used the Wilcoxon singed-rank test as a non-parametric alternative to t-test for matched pairs. The results are reported in Table~\ref{tab:mean_diff_test}, where we observe that most of the core developers have been in both states (\texttt{non-coding} and \texttt{inactive}) at least once. We observe a few cases of statistically significant mean difference at the 1\% level (after Bonferroni-Holm correction for multiple comparisons) between \texttt{non-coding} and \texttt{inactive} break duration, namely flutter, atom, jquery, crystal-lang, and MinecraftForge. 
To assess the magnitude of such differences, we computed the effect size as Cliff's~$\delta$, using the thresholds provided by \citep{hess2004effsize}.\footnote{We also assessed effect size by computing the Glass rank biserial correlation coefficient \citep{king2008grb} and obtained the same results.} We found a large effect size for atom, crystal-lang, and MinecraftForge.}

\begin{table}[tb]
\caption{Results of the Wilcoxon signed-rank (paired) tests to reveal mean differences between \texttt{non-coding} and \texttt{inactive} break durations for developers who have been in both states (results in \textbf{bold} are significant at the 1\% level after p-value correction and also have a large effect size).}
\centering
\footnotesize
\begin{tabular}{lrrrrll}
\hline
 & \multicolumn{3}{l}{\textbf{\# developers who have been}} & \multicolumn{1}{l}{\textbf{}}  & \multicolumn{1}{l}{\textbf{}}  \\
 \textbf{Organization} & \textbf{non-coding}   & \textbf{inactive}  & \textbf{both}  & \textbf{W} & \textbf{adjusted p} & \textbf{Cliff's $|\delta|$}\\ \hline
nodejs          & 112   & 98    & 98    & 502   & 0.3875  & 0.5146 \\
rails           & 92    & 93    & 92    & 168  & 1.0 & 0.6832 \\
laravel         & 72    & 71    & 70    & 132   & 0.3875 & 0.5527 \\
github          & 68    & 64    & 60    & 20 & 0.9375 & 0.7594  \\
facebook        & 33    & 33    & 33    & 45   & 1.0 & 0.5914 \\
flutter         & 27    & 24    & 23    & 73    & 1.5905e-10* & 0.2949\\
fastlane        & 17    & 17    & 17    & 7    & 1.0 & 0.6678  \\
jekyll          & 16    & 15    & 15    & 6    & 1.0 &  0.6533 \\
elixir-lang     & 13    & 13    & 13    & 12    & 0.1719 & 0.4142  \\ 
atom            & 10    & 11    & 10    & 8    & \textbf{1.2379e-09}* & \textbf{0.4800}$^\diamond$ \\
jquery          & 9     & 8     & 8     & 14    & 3.0262e-13* & 0.1250  \\
ionic-team      & 8     & 8     & 8     & 11     & 0.0103 & 0.1719 \\
BabylonJS       & 7     & 7     & 7     & 1     & 1.0 & 0.5306 \\
JabRef          & 8     & 7     & 7     & 0     & 0.1719 & 0.9592  \\
crystal-lang    & 6     & 6     & 6     & 6     & \textbf{0.0038}* & \textbf{0.6111}$^\diamond$  \\
MinecraftForge  & 6     & 6     & 6     & 3     & \textbf{7.2804e-10}* & \textbf{0.5556}$^\diamond$\\
aseprite        & 1     & 1     & 1     & N/A     & N/A & N/A  \\
SpaceVim        & 1     & 1     & 1     & N/A   & N/A & N/A  \\ \hline
*$p < 0.01$ \\
\multicolumn{7}{l}{$|\delta|<0.147$ "negligible", $|\delta|<0.330$ "small", $|\delta|<0.474$ "medium", otherwise $^\diamond$"large"}
\end{tabular}
\label{tab:mean_diff_test}
\end{table}

\vspace{3mm}
\begin{tcolorbox}[standard jigsaw,
	title=RQ3 -- Break length,
	opacityback=0]
	\small
	\begin{itemize}
		\item On average, \texttt{inactive} breaks last longer than \texttt{non-coding} breaks (65 vs. 32 days).
		\revisionone{\item This difference is, however, statistically significant only for three organizations (atom, crystal-lang, and MinecraftForge).}
	\end{itemize}
\end{tcolorbox}

\subsection{RQ4 -- State Transition Probabilities}
\label{sec:res_RQ4}

\begin{figure}[tb]
 \centering
\includegraphics[width=0.7\textwidth]{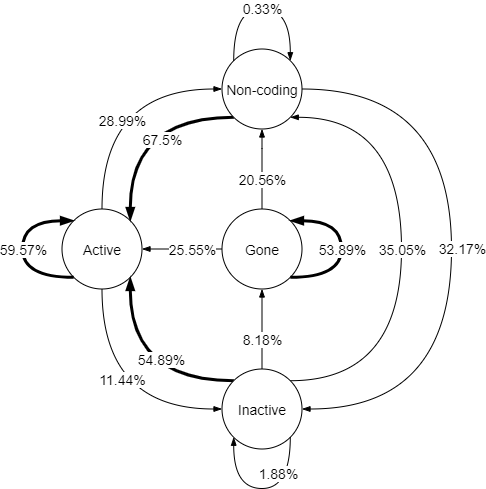}
\caption{Aggregated transition probabilities all core developers in the sampled organizations.}
\label{fig:cumulative_chain}
\end{figure}

\begin{figure*}
\setcounter{subfigure}{12}

\centering
    \subfloat[linguist\label{fig:linguist_chain}]{
       \includegraphics[width=0.5\textwidth]{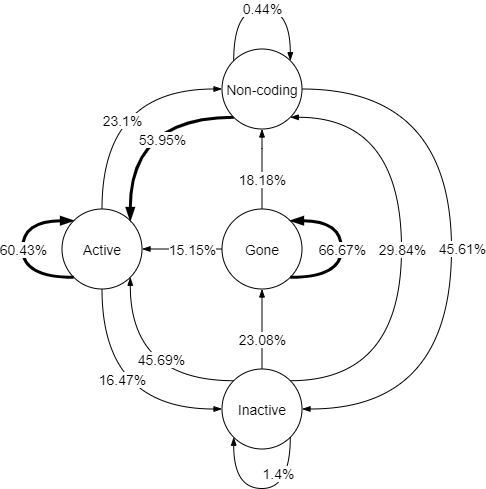}
     }
    \subfloat[atom\label{fig:atom_chain}]{%
       \includegraphics[width=0.5\textwidth]{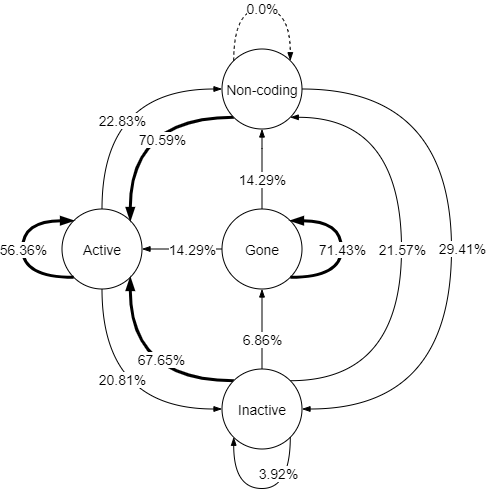}
     }
     \hfill
     \subfloat[ionic\label{fig:ionic_chain}]{%
       \includegraphics[width=0.5\textwidth]{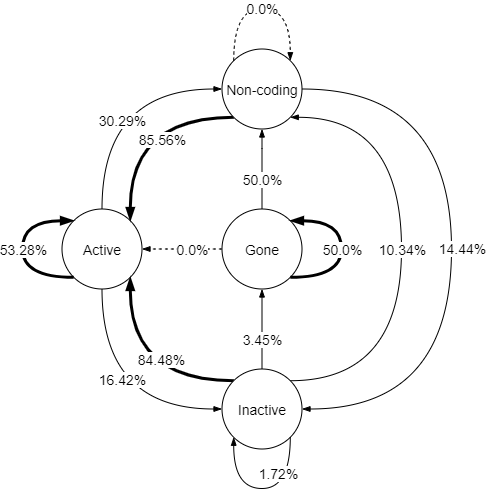}
     }
     \subfloat[react\label{fig:react_chain}]{%
       \includegraphics[width=0.5\textwidth]{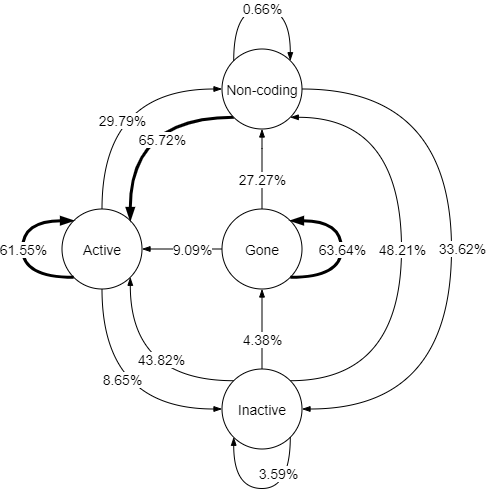}
     }
     \hfill
     \subfloat[jquery\label{fig:jquery_chain}]{%
       \includegraphics[width=0.5\textwidth]{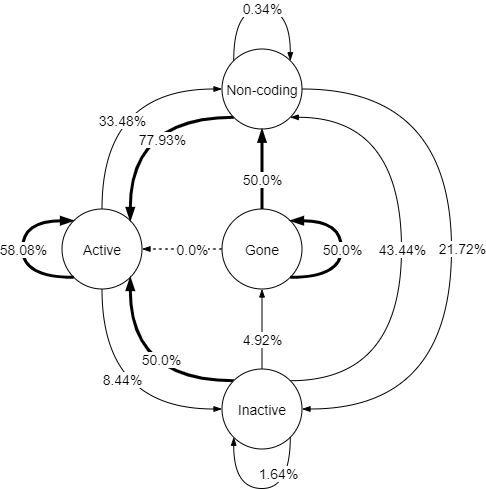}
     }
     \subfloat[flutter\label{fig:flutter_chain}]{%
       \includegraphics[width=0.5\textwidth]{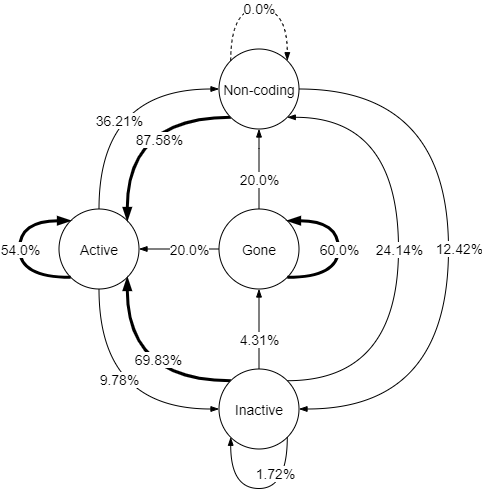}
     }
     \caption{Transition probabilities for core developers of the analyzed organizations.}
    \label{fig:markov_chains_company}
\end{figure*}

After identifying all the core developers' breaks, we calculated their probabilities to transition from one state into another. Specifically, in Fig.~\ref{fig:cumulative_chain}, we present the transition probabilities aggregated for all the sampled organizations, 
whereas in Fig.~\ref{fig:markov_chains_company} we report the probabilities for each organization individually.



\begin{figure*}
\setcounter{subfigure}{6}
\centering
    \subfloat[jabref\label{fig:jabref_chain}]{%
       \includegraphics[width=0.5\textwidth]{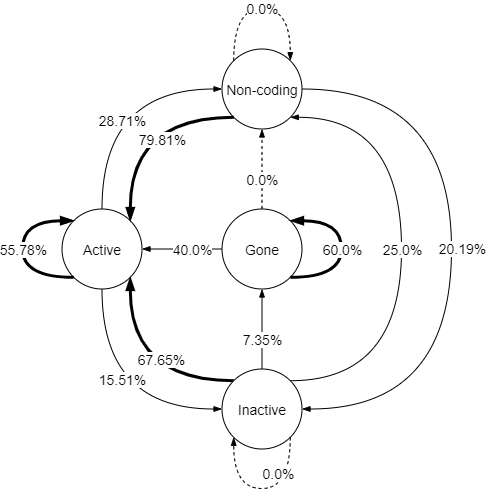}
     }
    \subfloat[SpaceVim\label{fig:SpaceVim_chain}]{%
       \includegraphics[width=0.5\textwidth]{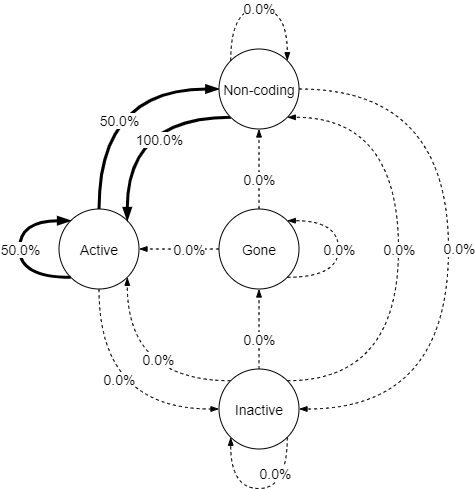}
     }
     \hfill
     \subfloat[fastlane\label{fig:fastlane_chain}]{%
       \includegraphics[width=0.5\textwidth]{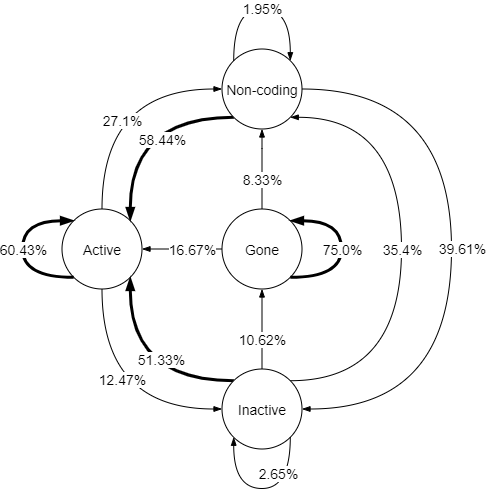}
     }
     \subfloat[crystal\label{fig:crystal_chain}]{%
       \includegraphics[width=0.5\textwidth]{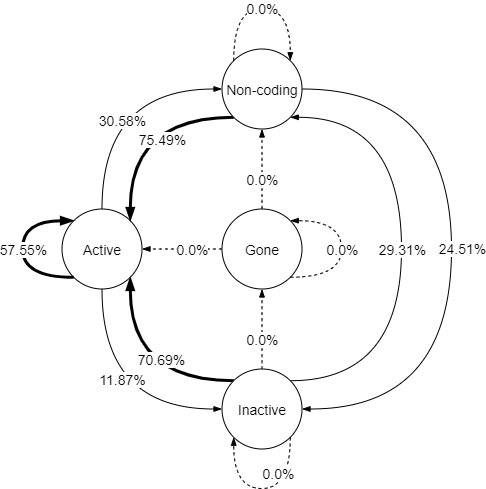}
     }
     \hfill
     \subfloat[BabylonJS\label{fig:Babylonjs_chain}]{%
       \includegraphics[width=0.5\textwidth]{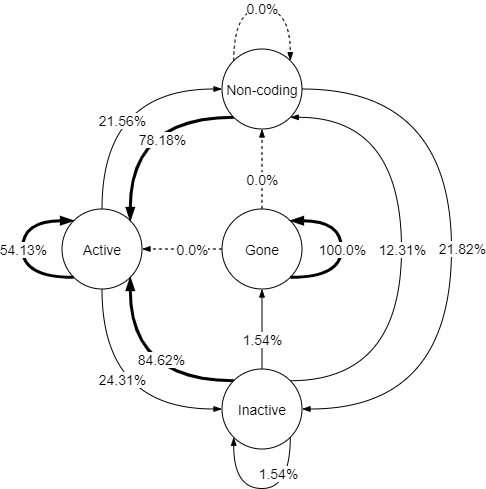}
     }
     \subfloat[elixir\label{fig:elixir_chain}]{%
       \includegraphics[width=0.5\textwidth]{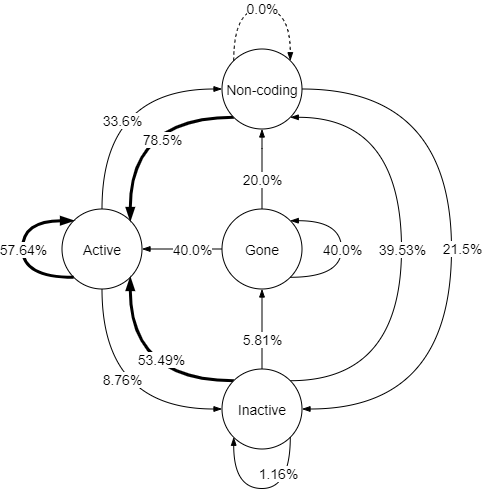}
     }
\end{figure*}

\noindent\textbf{Active state}. From Fig.~\ref{fig:cumulative_chain}, we can observe that the average probability to remain \texttt{active} is 59.57\%. This suggests that more than a half of the analyzed core developers tend to follow a constant rhythm of code contributions. 
This is similar for each organization in our sample (probabilities in the range 50.0--62.9\%, see Fig.~\ref{fig:markov_chains_company}). On average (see Fig.~\ref{fig:cumulative_chain}), core developers in the \texttt{active} state are more likely to transition to \texttt{non-coding} (28.99\%) than to become \texttt{inactive} (11.44\%). This observation also holds true when we analyzed organizations individually (Fig.~\ref{fig:markov_chains_company}), with the exception of BabylonJS (see Fig.~\ref{fig:Babylonjs_chain}).


\begin{figure*}
\centering
    \subfloat[node\label{fig:node_chain}]{%
       \includegraphics[width=0.5\textwidth]{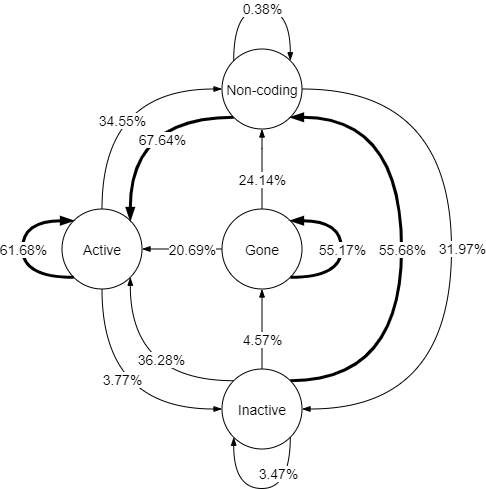}
     }
    \subfloat[rails\label{fig:rails_chain}]{%
       \includegraphics[width=0.5\textwidth]{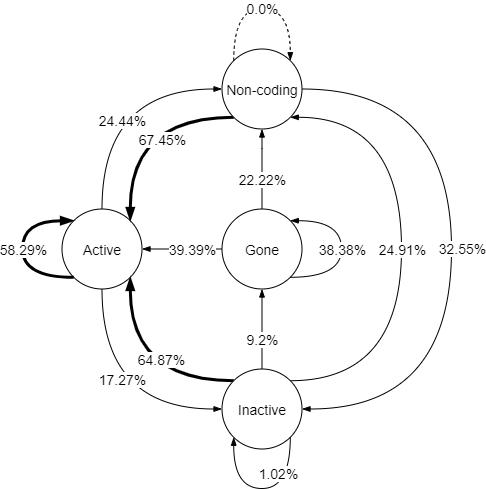}
     }
     \hfill
     \subfloat[aseprite\label{fig:aseprite_chain}]{%
       \includegraphics[width=0.5\textwidth]{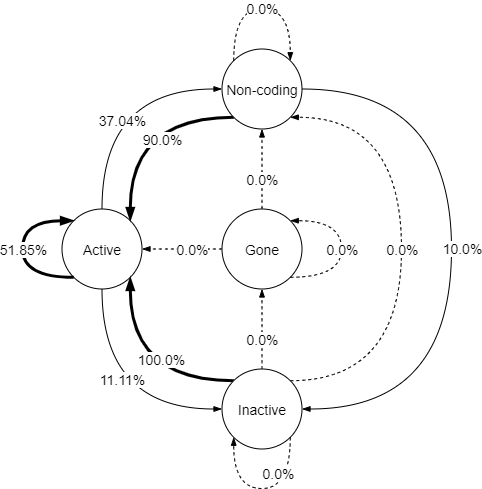}
     }
     \subfloat[jekyll\label{fig:jekyll_chain}]{%
       \includegraphics[width=0.5\textwidth]{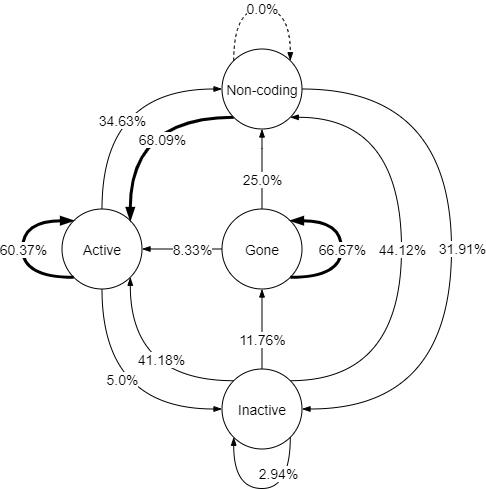}
     }
     \hfill
     \subfloat[laravel\label{fig:laravel_chain}]{%
       \includegraphics[width=0.5\textwidth]{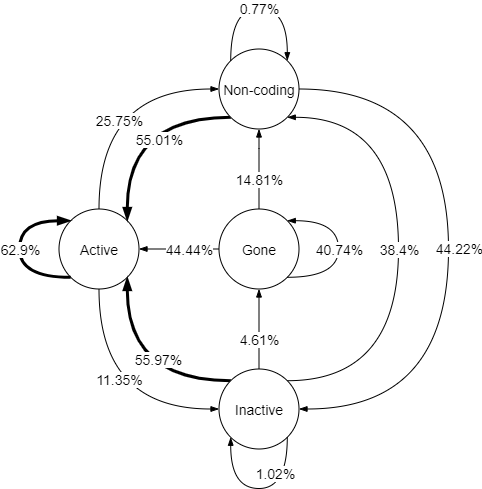}
     }
     \subfloat[MinecraftForge\label{fig:MinecrafForge_chain}]{%
       \includegraphics[width=0.5\textwidth]{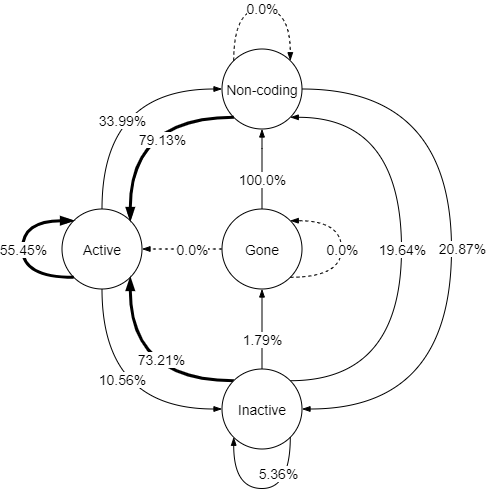}
     }
\end{figure*}

\noindent\textbf{Non-coding state}. On average, core developers in the \texttt{non-coding} state are much more likely to return to being \texttt{active} (67.5\%, see Fig.~\ref{fig:cumulative_chain}) than remain in the same state (0.33\%) or become \texttt{inactive} (32.17\%). 
The same observation also holds true at the individual organization level (Fig.~\ref{fig:markov_chains_company}).
 
\noindent\textbf{Inactive state}. We found that \texttt{inactive} core developers who do not provide any signal of life/participation in an organization are more likely to resume contributing by writing code---i.e., return to the \texttt{active} state (54.89\%, Fig.~\ref{fig:cumulative_chain})---than by collaborating without code---i.e., return to the \texttt{non-coding} state (35.05\%). Also, our findings suggest that \texttt{inactive} core developers have a lower probability (8.18\%) to remain so for more than one year and enter the \texttt{gone} state than to remain \texttt{inactive} for less than 12 months (1.88\%). Regarding the individual organizations, despite several  differences observed, no clear pattern can be identified. 

\noindent\textbf{Gone state}. On average, more than a half of the developers who abandon organizations for more than one year remain in the \texttt{gone} state (53.89\%). Still, \textit{comebacks} are not rare events, with 25.55\% and 20.56\% going back to the \texttt{active} and \texttt{non-coding} state, respectively. However, looking at the organizations individually, a quite varied picture emerges, which does not appear to be affected by their type. First, we notice that in only 4 out of 18 organizations (i.e., aseprite, MinecrafForge, SpaceVim, and crystal-lang), 
no core developer has ever disengaged, i.e., transitioned into the \texttt{gone} state. 
\revisionone{In a way, this pattern of behavior is expected for these smaller communities in which the main project relies on 1-3 Truck Factor developers and 1-6 core developers (see Table~\ref{tab:tfcomparison}), whose disengagement would jeopardize the survival of both the project and the community. As such, these developers form a solid base of project maintainers who have never taken breaks longer than a year. For the remaining organizations for which we found cases of disengagement, such as node (Fig.~\ref{fig:node_chain}) and linguist (Fig.~\ref{fig:linguist_chain}), we notice that nearly 60\% of them remain in the \texttt{gone} state (between 40-100\%), while the others resume contributing, i.e., go back to either \texttt{active} or \texttt{non-coding}.}

To further our understanding of developers disengaging from projects, we analyzed the association between the level of code contribution to a project and the likelihood of developers transitioning into the \texttt{gone} state at least once. To do so, we used the set of core developers ($n=538$) with their associated percentage of code contribution to the projects. Then, we computed the per-project median and assigned each developer to either of two equally-sized bins, called \textit{high}- and \textit{low}-level contributors. Finally, we built a logistic regression model where \textit{is\_gone} is the dichotomous dependent variable to be predicted (i.e., whether a developer has ever transitioned to the \texttt{gone} state at least once) and the only predictor is the ordinal variable \textit{contribution\_level} (low/high).

\begin{table}[b]
\caption{Results of the logistic regression shows that for high-level contributors the odds of disengaging at least once from projects are 61\% (OR=0.39) less than low-level contributors.}
\label{tab:logit}
\centering
\begin{tabular}{lccc}
\hline
\multirow{2}{*}{Core devs ($n=538$)}         & \multirow{2}{*}{\textbf{OR}}     & \multicolumn{2}{c}{\textbf{Confidence Interval}}         \\ \cline{3-4} 
                          &                                  & \textbf{2.5\%}             & \textbf{97.5\%}             \\ \hline
(Intercept)               & 1.276                            & 1.090                      & 1.618                       \\
contribution\_level = high     & 0.392                            & 0.275                      & 0.555                       \\ \hline
\multicolumn{4}{l}{AIC = 716.46}  
\end{tabular}
\end{table}

The results of the logistic regression are reported in Table~\ref{tab:logit}. The odds ratio (OR) for the variable \textit{contribution\_level = high} is 0.39. The 95\% confidence interval for the OR does not include 1, indicating that the result is statistically significant. Therefore, as compared to low-level contributors, high-level contributors' odds of disengaging from a project at least once (i.e., \textit{is\_gone = true}) decrease by 61\%.

\vspace{3mm}
\begin{tcolorbox}[standard jigsaw,	title=RQ4 -- State transition probabilities,
	opacityback=0]
	\small
	\begin{itemize}
		\item Most core developers have a steady rhythm of code contributions, as the average probability of remaining in the \texttt{active} state is \tildex{60}\%.
		\item Most core developers who pause their code contributions and transition into the \texttt{non-coding} and \texttt{inactive} states resume coding and go back to \texttt{active} (on average \tildex{67\%} and \tildex{55\%}, respectively).
		\item On average, about \tildex{8\%} of \texttt{inactive} developers remain so for over a year and eventually become \texttt{gone}, disengaging from the organization.
		\item On average, more than a half of \texttt{gone} developers never come back (\tildex{54\%}), whereas the remaining half is equally distributed between those who resume coding (i.e., \texttt{active}, \tildex{26\%}) and those who resume collaborating without contributing code (i.e., \texttt{non-coding}, \tildex{21\%}).
		\revisionone{ \item No transition into the \texttt{gone} state was observed for smaller projects, counting on 1-3 Truck Factor developers and up to 6 Core developers.} 
		\item As compared to Low-level contributors, for High-level contributors the odds of disengaging from a project at least once decrease by 61\%.
	\end{itemize}
\end{tcolorbox}

\section{Discussion}
\label{sec:discussion}


In the following, we discuss our results, presenting implications, insights, and future research avenues. 

\noindent\textbf{Rhythm-based model}. Our revised model, presented in Fig.~\ref{fig:statemodel}, is a significant extension of the original model presented in~\citep{soheal2019}---inspired by the circadian rhythm, which also varies from individual to individual. The assessment of the revised model by the developers showed that 94\% of them agreed with it, and that it correctly captured about 70-80\% of the transitions (as per RQ1, see Sect.~\ref{sec:res_RQ1}). Notably, some developers did not remember their breaks or did not acknowledge the transitions. 
For many cases of transition disagreements, we found that the developers were not actually working on the project, but they were still present in some form (e.g., lurking, waiting for the next thing to do). Although the developers considered themselves active in the project, per the model definition they fit the inactivity state, since they were considered not actively contributing. We also found that even considering multiple events and activities that are public, contributors perform other kinds of activities not observable from the traces left on \textsc{GitHub}. This corroborates the results from \cite{trinkenreich2020hidden}, who provide empirical evidence of the existence and importance of community-centric roles (e.g., advocate, license manager, community founder), which typically remain hidden, since these roles may not leave traces in the software repositories typically analyzed by researchers. 


\noindent\textbf{Why are you leaving?} Our investigation (RQ1) collected evidence on the reasons why core developers who do not completely disengage take breaks and then resume their activity. Albeit not exhaustive, our findings show that in both cases the reasons are mostly personal (\tildex{80\%}) rather than related to the community (e.g., social interactions, technical changes, and organizational aspects). Such personal reasons concern life events (e.g., child birth, vacation) as well as professional (e.g., new job) and financial changes (e.g., stopped volunteer contribution to focus on a paid job). 


\revisionone{ These reasons for disengagement extend the current literature that focuses on motivations to join and factors that explain longer attachment~\citep{zhou2012, silva2020google, lin2017developer, constantinou2017, gerosa2021motivation}. We contribute by showing that, in addition to job-related reasons (also reported also by \cite{Miller2019}),  a non-negligible share of the reasons to leave are related to life-events and change of interest. These two latter categories of reasons cannot be easily controlled by project maintainers and are harder to predict based on historical data. On the other hand, we found cases of disengagement related to changes in the community organizational structure, problem with other members' ideas, and lack of available tasks. These reasons are in line with other studies, which showed that changes in the organization~\citep{yu2012empirical} and mismatch of ideas~\citep{steinmacher2018almost} are reasons for contributors to give up. Still, we show that those core members who leave for job-related and life-event reasons generally return---which explains more than 50\% of probability of contributors moving away from the \texttt{gone} state (see Fig.~\ref{fig:cumulative_chain}).
}


\noindent\textbf{Variations in inactive periods.} Our findings serve as an input to communities that want to create awareness mechanisms to inform about developers' inactivity. As our results showed, almost all developers take some kind of break during their life cycle---89\% and 97\% have been respectively in the \texttt{inactive} and \texttt{non-coding} states at least once (as per RQ2, see Sect.~\ref{sec:res_RQ2}). Moreover, as investigated in RQ3, the duration of the breaks may vary widely, even across developers from a single project (see Fig.\ref{fig:sleep_hiber_durations}). We noted, for example, that for rails/rails, the period in \texttt{non-coding} state varies from 8 to 270 days, while for atom/atom, the variation in \texttt{gone} state is 428-651. This suggests that projects and organizations may find reference values based on the distribution of breaks; however, it is important to closely follow the individual rhythm and characteristics of developers, given that each developer may have a distinct contribution pace.


\noindent\textbf{It is not a `goodbye,' it is a `see you soon.'} One interesting finding from RQ2 is that about one third of the core developers analyzed (33\%) have been \texttt{gone} at least once from a project for one year or longer (see Table~\ref{tab:inactive_cbh_developers}). Furthermore, as per RQ4 (see Sect.~\ref{sec:res_RQ4}), the transition probabilities presented in Fig.~\ref{fig:cumulative_chain} show that the probability of a \texttt{comeback}, considering all projects, is 25.6\%. Moreover, the chances that someone \texttt{inactive} will transition to \texttt{gone} is on average 8.2\%, which is much smaller than the chances of transitioning from \texttt{inactive} to \texttt{active} (54.9\%) or \texttt{non-coding} (30.1\%). 

Overall, the number of break events observed and the odds in Figs.~\ref{fig:cumulative_chain}--\ref{fig:markov_chains_company} suggest that the core members of the analyzed projects are more likely to stay than to abandon projects\revisionone{, especially for smaller projects like aseprite, SpaceVim, MinecraftForge, and crystal, which count on a smaller core base (see Table~\ref{tab:tfcomparison})}. Consistently, the results of the logistic regression reported in Table~\ref{tab:logit} indicate that projects remain sustainable because people with high-level contribution rates have higher odds (61\%) to stay active. This shows that turnover at the core-developers level is not an issue as we expected---considering the analyzed sample. Moreover, given the age of the projects analyzed (3 to 15 years, median 8), some turnover is expected with the rise of new core members. For example, as reported in Sect.~\ref{sec:res_RQ2} (see Table~\ref{tab:inactive_cbh_developers}), 38 core developers from rails were \texttt{gone} at the time of the data collection (41\% of the rails sample), and likewise for fastlane (9 devs., 47\%) and jekyll (8 devs., 47\%). The remaining core developers were, in general, enough to keep the projects running. Although the renewal was not analyzed in this paper, this is something worth investigating from the perspective of project sustainability. Future work may focus on the impact of breaks and disengagement with the project performance, and in understanding how other developers take over when a core member transitions into \texttt{gone}. 

\revisiontwo{
\textbf{Non-coding activities matter.} Usually, the indicators of contribution (such as GitHub graphs)\footnote{e.g., \url{https://github.com/rails/rails/graphs/contributors}} focus on contributions made to the repository, which represent mainly code contributions. However, it is known that contributors follow different pathways to contribute~\citep{trinkenreich2020hidden} and non-code contributions are often neglected~\citep{ferreira2017comparison, decan2020gap, Miller2019}. We found in our analysis that almost all core contributors take breaks from code contributions but take over non-coding roles (see Table~\ref{tab:inactive_cbh_developers}). Separating the analysis of these two types of contributions and making non-coding one state in the model is a way of not obfuscating the importance of---and give visibility to---non-code contributions.}

\begin{figure}[t]
\centerline{
\includegraphics[width=.75\textwidth]{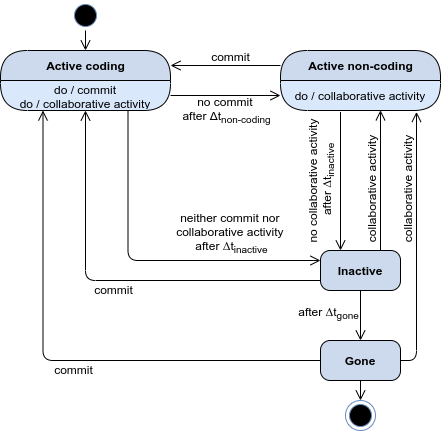}}
\caption{Final version of the inactivity model with revised names for \texttt{active-coding} and \texttt{active non-coding} states.}
\label{fig:finalstatemodel}
\end{figure}

\noindent\textbf{Adjustments to the model.} Given the feedback received during the qualitative assessment, we accordingly adjusted the model. First, we had to account for the `degrees of activity' provided by our model. Our analysis of RQ1 (see Sect.~\ref{sec:res_RQ1}) revealed that the model was accepted by 94\% of the surveyed developers. However, we noticed room for improvement based on the received feedback. One shortcoming that clearly emerged from the analysis of comments was that the \texttt{activity} state came with different degrees of action and that the model seemed to convey the idea that committing source code is the ultimate form of `action.' While it is impossible to represent all the nuances of development activities in a state diagram, we accepted the feedback and refined the model so that it would not unintentionally mislead developers into thinking that \texttt{non-coding} means that they are not actively participating. As such, in the final version of the model, presented in Fig.~\ref{fig:finalstatemodel}, we renamed the \texttt{active} state to \texttt{active coding} and \texttt{non-coding} to \texttt{active non-coding}.

Regarding the identification of the transitions into the \texttt{gone} state, the feedback we received showed that 4 out of 5 developers did not acknowledge that they had been \texttt{gone} (see Sect.~\ref{sec:res_RQ1}). As aforementioned, these developers point out that they \textit{did not want} to leave; rather, they state that they did not find anything to work on (D-04) or were too busy with their paid job (D-05). We argue that the disagreement is related to different perspectives taken: project vs. personal. Our model takes the perspective of the project, mapping when someone takes an overly long break. When a developer reaches the threshold of one year without any traces, per the model definition they are considered \texttt{gone}. 


\noindent\textbf{Future Work Opportunities and Implications for Researchers.} 
\revisionone{ We designed and evaluated our model and method to understand the different states of the contributor's life cycle and their transitions. Researchers can extend our approach to build prediction models to anticipate breaks and disengagement. An interesting future work would be to create and evaluate such prediction tools by extending the work of~\cite{decan2020gap} to consider the length of the breaks and \texttt{non-coding} activities as part of developers' contributions.

Researchers can also expand our research by including maintainers' perception of core developers taking breaks. We have so far focused on identifying and acknowledging breaks according to developers' own perception, but we do not yet know how these inactivity periods and breaks are seen by other community members.

Although we mined multiple public events from GitHub, sometimes we considered developers as not active because their activities did not leave traces on GitHub. Researchers who investigate contributors' activity in software repositories should be aware of this kind of activity. Usually, the literature focuses on coding activities, disregarding other kinds of contributions and `hidden figures'~\citep{trinkenreich2020hidden}. Although we only looked for the activities available on GitHub, this opens doors for further research about `invisible collaboration,' which may not be mined from project repositories.

A larger sample of repositories can also be investigated, including some private repositories owned by companies, to gather a better picture of how disengagement varies across different projects and ecosystems, and whether developers migrate. In addition, we did not analyze whether the breaks are related to the project schedule or coordinated among team members. 
A potential future direction would be an in-depth exploration of these relationships. }

\revisiontwo{Another way to expand our work is to look at the effects of sponsorship on developers' inactivities---to identify potential differences in frequency and length for company-supported OSS projects that routinely receive contributions from employed developers---as well as study the impact on productivity at the project level---to understand whether and how overall productivity takes a hit during core contributors’ breaks.}

\noindent\textbf{Implications for Practitioners.} \revisionone{ Our results show that developers have different work rhythms, and that the transitions to breaks and abandonment vary from project to project. In fact, this is in line with the literature that has shown software engineering metrics to be highly context dependent~\citep{zhang2013does, gil2016software, aniche2016satt}. Nevertheless, in this paper we presented individual analyses of a variety of projects and organizations, which allows practitioners to observe similarities and differences, fostering the generalization by similarity~\citep{wieringa2015six}. Our results also indicate a range of values to which projects can compare themselves.

Inspired by our model and methods, communities can also build tools to monitor the rhythm of individual developers (e.g., building dashboards) and detect when they take longer-than-usual breaks, helping maintainers and managers to assess developers' life cycles individually. While informally discussing our results with a core developer who participated in the research, he mentioned that this kind of information would be valuable to create a ``\textit{Contributor Relationship Management System}'' (CRM). In this case, according to him, ``\textit{customers are the contributors. As an OSS project owner, one wants to keep the contributors contributing and identify high-potential ones. When I see high-potential contributors being away longer than usual, I would use the communication channel they like}.'' Such a system could work as a heart monitor, alerting the community managers of potential cases of `non-reported' breaks. This may help maintainers, for example, identify developers who are `waiting for the next thing to do,' and have been inactive for longer than expected. It is important to keep these developers in the loop, because they may eventually lose track of the technical and social aspects of the project, which may lead to disengagement. This is especially important since most of the reasons to leave reported by our survey respondents are personal (see Table~\ref{tab:codingbreaks}), beyond the project's reach.
}

\noindent\textbf{Limitations}. Every empirical study suffers from threats to validity and limitations~\citep{Wohlin-2012}. First, we acknowledge that we focused on studying a sample of eighteen projects and organizations hosted on \textsc{GitHub}.\revisiontwo{Although  GitHub hosts millions of open source projects~\footnote{\url{https://octoverse.github.com/\#the-world-of-open-source}}, we acknowledge that the results may be different for other forges and ecosystems.  
Future research can expand our work to consider other environments. Moreover,} \revisionone{because we used a convenience sampling strategy, we acknowledge the potential sampling bias. Also, despite our efforts to add variety to the sample, most of the projects analyzed are popular development frameworks and tools for developers---with the exceptions of jabref, MinecraftForge, aseprite, and SpaceVim. This imbalance is likely the side effect of having started the sampling process by selecting projects and organizations from those trending on \textsc{GitHub}. Accordingly, our findings may not generalize to other OSS communities, given that each project and organization has their peculiarities. Nonetheless, the projects in our sample are diverse in terms of programming languages, size, age, and popularity, which helped us provide an initial understanding of the developers' inactivity phenomenon. Presenting individual analyses of these projects and organizations allows the identification of similarities and differences, which fosters the generalization by similarity \citep{wieringa2015six}.} 



\revisiontwo{Also related to the generalization of results, after sampling the 18 organizations, in Sect.~\ref{sec:data-collection} we described how we selected their main project as the largest one in terms of the number of contributors, stars, forks received, pull requests, and LOC. While most of the sampled projects are undoubtedly representative of the organizations---since they have identical or closely resembling names (e.g., atom/atom)---this might not be the case for github/linguist and facebook/react. Both github and facebook organizations are `umbrellas' hosting any open-source effort from the two companies and, as such, they include a varied set of loosely related projects. Hence, we acknowledge that linguist and react may not be representative of their respective organization and that this difference may impact the generalizability of our findings.}

The method designed to identify breaks showed a reasonably good performance (between 70-80\% of developers acknowledged their state and transitions), albeit not perfect. One possible explanation for the disagreements is that we consider only the activities tracked by \textsc{GitHub}, whereas other actions happen outside of the platform, like communicating over email or \textsc{Gitter}. While we acknowledge this as a limitation of our current method, we point out that there is a great variety of communications solutions (not to mention the other non-communication tools), which makes it impracticable to cover all of possible alternatives. In addition, these tools do not always leave freely accessible traces of activity. Projects using \textsc{Slack}, for example, make it particularly hard to download and analyze the communication logs. Moreover, when integrating data from multiple sources, we point out the need to disambiguate developers' multiple logins and identifications~\citep{wiese2016mailing}. Finally, we hope that, by making all the scripts available, there can be independent replications and extensions of our work that bridge this gap by investigating the effect of external communication channels on our method. 



When identifying the \texttt{gone} breaks, our model cannot distinguish between cases in which developers have the \textit{intention of abandoning} a project from those who are just taking a break longer than a year (for whatever reason) until they will eventually come back and resume their activity. \revisionone{ Furthermore, the survey about the reasons for taking breaks and resuming contribution (described in Sect.~\ref{sec:res_RQ1}) has revealed that reasons can also depend on the decision of the employer. It is indeed more and more common for paid developers to contribute to OSS projects; some companies even share their products under open-source licenses, while keeping their employees in charge of maintaining the projects \citep{pinto2018challenges}. In our current analysis, we do not distinguish between company-sponsored projects, which routinely receive contributions from employed developers, and community-based projects, which mostly grow through voluntary contributions. In fact, the distinction between the two types can be blurry, since there are often sponsored contributions to community-based projects and vice-versa. To date, there have been semi-automatic attempts at classifying the contribution between paid and non-paid developers \citep{pinto2018challenges}; however, there is no broadly accepted method to be employed at scale, without time-consuming manual intervention.}

Another potential limitation of this work relates to the choice of the algorithms used to identify the sample of core developers under investigation. After manual validation by comparing the Truck Factor developers sample with the contributors listed in the projects' contributors pages on \textsc{GitHub},\footnote{For example: \url{https://github.com/ionic-team/ionic/graphs/contributors}} we observed that the top contributor was not automatically selected in the cases of jabref and ionic-framework, and we had to manually amend these errors. As such, the Truck Factor algorithm may suffer from shortcomings in identifying all the core developers, arguably due to how authorship of a file is determined. However, by also using the Commit-Based Heuristics approach, which broadens the list of core developers of each project to the authors of 80\% of non doc-only commits, we implicitly address those limitations.

Lastly, the method proposed to analyze the developer's life cycle may be seen as arbitrary. To mitigate this issue, we evaluated our approach via survey twice, first when conceiving the model (Sect.~\ref{sec:phase_one}), and then as an answer to RQ1 (Sect.~\ref{sec:res_RQ1}), which resulted in the final version depicted in Fig.~\ref{fig:finalstatemodel}.

\section{Conclusion}
\label{sec:conclusions}

In this paper, we analyze the life cycle of OSS project developers concerning their breaks and disengagement. First, by collecting data from \textsc{GitHub} and conducting a survey with developers, we devised a model that describes states and transitions related to the breaks. Then, we designed and implemented an approach to identify developers' transitions to/from inactivity states, which we validated by collecting feedback from developers---94\% of the TF core developers agree with our state model of inactivity and \tildex{71\%} and \tildex{79\%} of them acknowledge their breaks and state transitions, respectively. 

Our results also showed that breaks are rather common, and that core members take frequent breaks, which vary in length and type. We observed that all developers took at least one break, 97\% of them transitioned to \texttt{non-coding} and 89\% to \texttt{inactive}. 

We also analyzed the probability of the transitions to/from inactivity states and observed that core developers will likely remain in the projects. However, if they transition to \texttt{gone}, they are less likely to come back (\tildex{54\%} of chance, on average). Our results may help communities to create mechanisms to monitor core members breaks or to prepare to recruit new members in case of disengagements.

\begin{acknowledgements}
We thank all the developers who participated in the interviews and surveys. This research is partially funded by NSF grants IIS-1815503 and IIS-1900903, and by the Brazilian National Council for Scientific and Technological Development (CNPq) grant \#313067/2020-1. The computational work has been executed on the IT resources made available by two projects, ReCaS and PRISMA, funded by MIUR under the program PON R\&C 2007-2013.
\end{acknowledgements}

\bibliographystyle{spbasic} 
\bibliography{references}

\newpage
\appendix

\section{Breaks Algorithm}\label{appendix:algo}
This is the pseudocode of our sliding-window algorithm for identifying breaks in the commit history of developers.

A window $W$ is defined as a couple $<start, end>$. A pause $p$ is defined as a couple $<start, end>$ in between two consecutive activities. A break $b$, which by definition is a pause in a window $W$ that is larger than the far out value threshold $t$, is therefore defined as a tuple $<W, p, t>$. To compute the far out value threshold, a window $W$ must contain at least 4 pauses (data points); if so, $W$ is considered valid. Finally, a threshold $t$ in a window containing a distribution of pauses for which the $IQR \leq 1$ is considered not valid. IQR, defined as the range holding 50\% of the data, is a measure of variability. Small values of IQR in a window suggest that the lengths of pauses concentrate around the mean value.

\newpage

\begin{algorithm}[H]
\SetAlgoLined
\KwData{$D$ \tcp*{the list of developers with at least one pause}
$win\_size$ \tcp*{the size of the window (def: 3 months)} 
$shift$ \tcp*{the window forward shift (def: 1 week)}
}
\KwResult{A list of breaks for all developers in $D$}
$All\_Breaks \leftarrow List()$\;
 \ForEach{$dev \in Devs$}{
  $life\_span \leftarrow Life(dev)$ \tcp*{\# days between 1st and last commit}
  $W \leftarrow GetFirstWindow(life\_span, win\_size)$\;
  $i \leftarrow 1$\;
  $Win\_Thresholds \leftarrow List()$\;
  $Longer\_Breaks \leftarrow List()$\;
  \While{$W \ne NULL$}{ 
   $Win\_Pauses \leftarrow GetPausesInWindow(W)$\;
   \tcp{Pauses that start in W but end after it}
   $Partially\_Included \leftarrow GetPausesPartiallyInWindow(W)$\;
   \tcp{if W is valid }
    \If{$Length(Win\_Pauses) \geq 4$}{
        $t\_fov \leftarrow ComputeFarOutValueThreshold(Win\_Pauses)$\;
        \tcp{Far out value threshold valid if IQR > 1}
        $valid \leftarrow IsValid(t\_fov)$\; 
        $Win\_Thresholds.Add(t\_fov, valid, i)$\;
    }
    \If{$Length(Win\_Pauses < 4)$ or $valid = FALSE$}{
        $prev\_t\_fov \leftarrow Win\_Thresholds.Get(i-1)$\;
        $prev\_valid = IsValid(prev\_t\_fov)$\;
        \If{$prev\_t\_fov \ne NULL$ and $prev\_valid = TRUE$}{
            $Win\_Thresholds.Add(prev\_t\_fov, prev\_valid, i)$\;
        }
        \Else{
            \tcp{pauses shorter than shift are safely ignored (not breaks)}
            \tcp{pauses longer than shift will be analyzed in the next W}
            $Win\_Pauses \leftarrow List()$\;
            \ForEach{$p \in Partially\_Included$}{
                \If{$Length(p) > win\_size$}{
                    \tcp{partially-included pauses longer than the window}
                    \tcp{size are breaks by definition}
                    $Longer\_Breaks.Append(p)$\;
                }
            }
            $W = NextWindow(life\_span, W, shift)$\;
            \textbf{continue}
        }
    }
     $Dev\_Breaks \leftarrow List()$\;
     \ForEach{$p \in Win\_Pauses \cup  Partially\_Included$}{
        \If{$Length(p) > t\_fov$}{
            $Dev\_Breaks.Append(<W, p, t\_fov>)$\;
        }
     }
     $W = NextWindow(W, shift)$\;
     \textbf{continue}
   }
   $avg\_t\_fov \leftarrow Mean(Win\_Thresholds)$\;
   \ForEach{$p \in Longer\_Breaks $}{
        $Dev\_Breaks.Append(<W, p, avg\_t\_fov>)$\;
   }
   $All\_Breaks \leftarrow All\_Breaks \cup Dev\_Breaks $\;
 }

\caption{Developers' breaks identification}
\end{algorithm}

\end{document}